\documentclass[a4paper,12pt]{article}
\usepackage{amsmath}
\usepackage{amsfonts}
\usepackage{amssymb}
\usepackage{graphicx}
\usepackage{subfigure}

\usepackage{color}
\usepackage[english]{babel}

\renewcommand{\baselinestretch}{1.667}

\newcommand{\cvec}{\mbox{\bf c}}

\newcommand{\xvec}{\mbox{\bf x}}

\newcommand{\vvec}{\mbox{\bf v}}
\newcommand{\zvec}{\mbox{\bf z}}
\newcommand{\Zvec}{\mbox{\bf Z}}

\newcommand{\Xvec}{\mbox{\bf X}}

\newcommand{\yvec}{\mbox{\bf y}}

\newcommand{\tvec}{\mbox{\bf t}}

\newcommand{\disp}{\displaystyle}
\newcommand{\sigmat}{\mbox{\boldmath $\Sigma$}}
\newcommand{\muvec}{\mbox{\boldmath $\mu$}}

\newcommand{\xivec}{\mbox{\boldmath $\xi$}}

\begin{document}

\title{\LARGE{\bf Multi-scale Classification using Localized Spatial Depth}} 
%\vspace{-2.5in}
%\\with Multiple Scales of Localization}}
\author{}
%\date{}
\maketitle

\begin{center}
 \large{Subhajit Dutta$^{a}$ and Anil K. Ghosh$^{b}$}

 \vspace{0.05in}
 \small
 {$^{a}$Department of Mathematics and Statistics,
 Indian Institute of Technology, Kanpur 208016, U.P., India.}\\
 {$^{b}$Theoretical Statistics and Mathematics Unit, Indian Statistical Institute,
 203, B. T. Road, Kolkata 700108, India.}\\

 \vspace{0.05in}
 {Email: tijahbus@gmail.com$^{a}$, akghosh@isical.ac.in$^{b}$}
\end{center}

{\renewcommand{\baselinestretch}{1.2}

\begin{abstract}
In this article, we develop and investigate a new classifier based on features extracted using spatial depth. Our construction is based on fitting a generalized additive model to the posterior probabilities of the different competing classes.
%Most of the existing depth based classifiers need the ellipticity of the population distributions to perform well.
To cope with possible multi-modal as well as non-elliptic %nature of the
population distributions, we develop a localized version of spatial depth and use that with varying degrees of localization to build the classifier. Final classification is done by aggregating several posterior probability estimates each of which is obtained using localized spatial depth with a fixed scale of localization. The proposed classifier can be conveniently used even when the dimension is larger than the sample size, and its good discriminatory power for such data has been established using theoretical as well as numerical results.
%for high-dimensional data,
%Further, this classifier can adequately deal with very general classification
%problems including situations, where the class distributions are non-elliptic having multiple modes. Using simulated and real benchmark data sets, the proposed classifier is shown to have competitive performance when compared with several well-known and widely used parametric and nonparametric classifiers. \\
%like those based on $k$-nearest neighbors, kernel density estimates, support vector machines, classification trees, random forests, artificial neural nets, etc.

% \begin{keywords}
\vspace{0.25in}
\end{abstract}
}

\newpage
\section{Introduction}

In a classification problem with $J$ classes, we usually have $n_j$ labeled observations $\xvec_{j1}, \ldots, \xvec_{jn_j}$ from the $j$-th class ($1 \leq j \leq J$), and we use these $n=\sum_{j=1}^{J} n_j$ observations to construct a decision rule for classifying a new unlabeled observation $\xvec$ to one of these $J$ pre-defined classes. If $\pi_j$ and $f_j$ respectively denote the prior probability and the probability density function of the $j$-th class, and $p(j|\xvec)$ denotes the corresponding posterior probability, the optimal {\it Bayes classifier} assigns $\xvec$ to the class  %having the maximum posterior probability. So, an observation $\xvec$ is classified to the class
$j^{*}$, where
$j^{*} = \arg \max_{1 \leq j \leq J} p(j|\xvec) = \arg \max_{1 \leq j \leq J} \pi_j f_j(\xvec)$.
%where $p(j|\xvec)$ is the posterior probability of the $j$-th ($1 \leq j \leq J$) class.
%It is known that this classifier has the lowest misclassification probability.
However, the $f_j(\xvec)$'s (or, the $p(j|\xvec)$'s) are unknown in practice, and one needs to estimate them from the training sample of labeled observations. Popular parametric classifiers like linear discriminant analysis (LDA) and quadratic discriminant analysis (QDA) are motivated by parametric model assumptions on the underlying class distributions. So, they may lead to poor classification when the model assumptions fail to hold, and the class boundaries of the Bayes classifier have complex geometry. On the other hand, nonparametric classifiers like those based on $k$-nearest neighbors ($k$-NN) and kernel density estimates (KDE) are more flexible and free from such model assumptions. But, they suffer from the curse of dimensionality and are often not suitable for high-dimensional data.

Consider two examples denoted by {\bf E1} and {\bf E2}, respectively. {\bf E1} involves a classification problem with two classes in $\mathbb{R}^d$, where the distribution of the first class is an equal mixture of N$_d({\bf 0}_d, {\bf I}_d)$ and N$_d({\bf 0}_d, 10{\bf I}_d)$, and that for the second class is N$_d({\bf 0}_d, 5{\bf I}_d)$. Here N$_d$ denotes the $d$-variate normal distribution, ${\bf 0}_d=(0, \ldots, 0)^{T} \in \mathbb{R}^d$ and ${\bf I}_d$ is the $d \times d$ identity matrix. In {\bf E2}, each class distribution is an equal mixture of two uniform distributions. The distribution for the first (respectively, the second) class is a mixture of U$_d(0,1)$ and U$_d(2,3)$ (respectively, U$_d(1,2)$ and U$_d(3,4)$). Here U$_d(r_1,r_2)$ denotes the uniform distribution over the region $\{\xvec \in \mathbb{R}^d : r_1 \leq \|\xvec\| \leq r_2\}$ with $0 \leq r_1 < r_2$.
Figure 1 shows the class boundaries of the Bayes classifier for these two examples when $d=2$, and $\pi_1=\pi_2=1/2$. The regions colored grey (respectively, black) correspond to observations classified to the first (respectively, the second) class by the Bayes classifier. It is clear that classifiers like LDA and QDA or any other classifier with linear or quadratic class boundaries will deviate significantly from the {Bayes classifier} in both examples. A natural question then is how standard nonparametric classifiers like those based on $k$-NN and KDE perform in such examples. In Figure 2, we have plotted average misclassification rates of these two classifiers along with the Bayes risks for different values of $d$. These classifiers were trained on a sample of size 100 generated from each class distribution, and the misclassification rates were computed based on a sample of size 250 from each class. This procedure was repeated 500 times to calculate the average misclassification rate. Smoothing parameters associated with $k$-NN and KDE (i.e., the $k$ in $k$-NN and the bandwidth in KDE) were chosen by minimizing leave-one-out cross-validation estimates of misclassification rates \cite{HTF09}. Figure 2 shows that in {\bf E1}, the Bayes risk decreases to zero as $d$ grows. Since the class distributions in {\bf E2} have disjoint supports, the Bayes risk is zero irrespective of the value of $d$. But in both examples, the misclassification rates of these two nonparametric classifiers increased to almost 50\% as $d$ increased.

%\vspace{-.5in}
\begin{figure}
\begin{center}
\vspace{-0.5in}
\subfigure[\vspace{-0.2in}Example {\bf E1}]{\includegraphics[width=4.5cm]{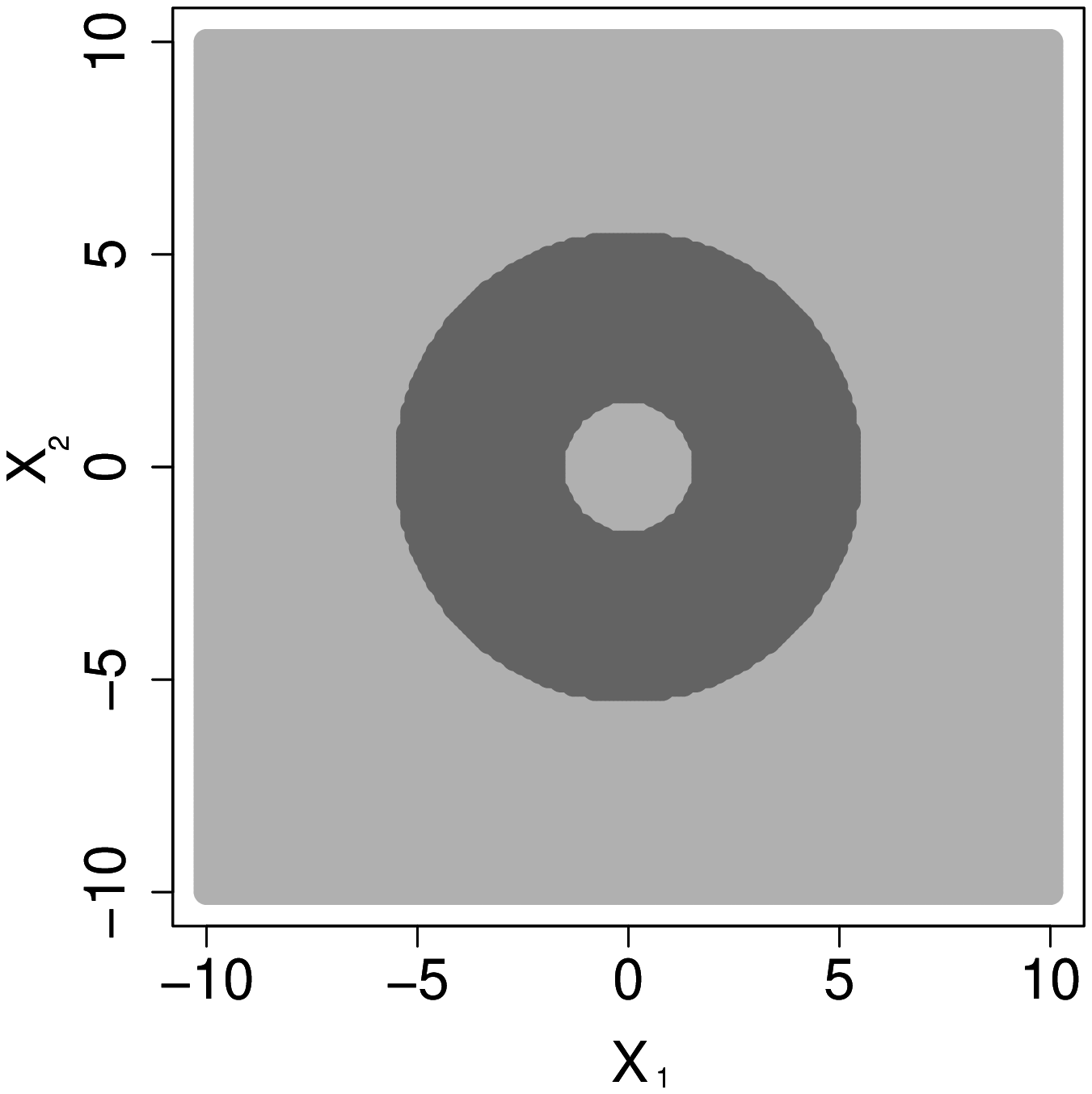}} \hfil
\subfigure[\vspace{-0.2in}Example {\bf E2}]{\includegraphics[width=4.5cm]{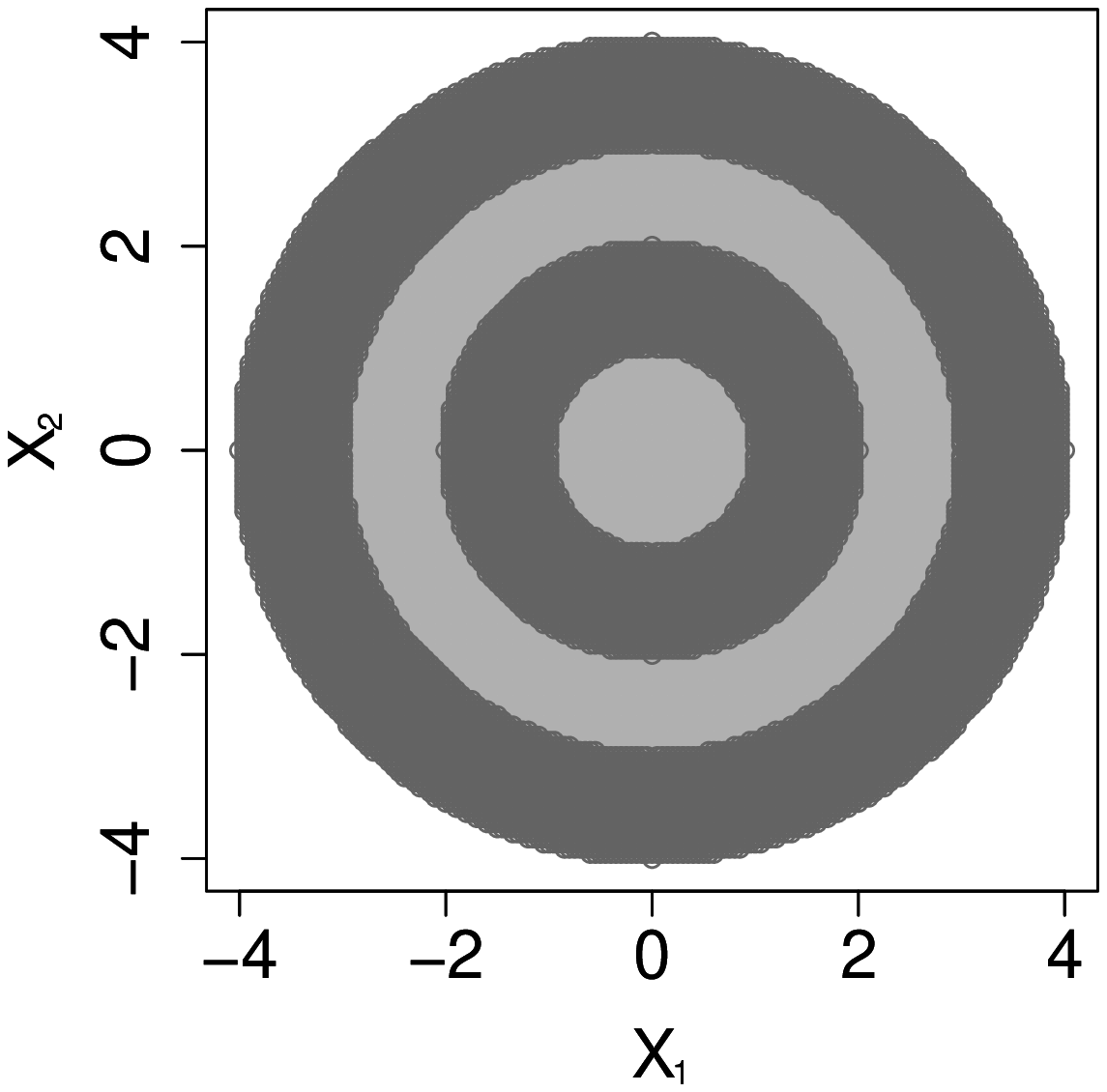}}
%\vspace{-0.5in}
\end{center}
\vspace{-.15in}
\caption{\tt Bayes class boundaries in $\mathbb{R}^2$.}
%\vspace{-.1 in}
\end{figure}

\begin{figure}
\begin{center}
\subfigure[Example {\bf E1}]{\includegraphics[width=5cm]{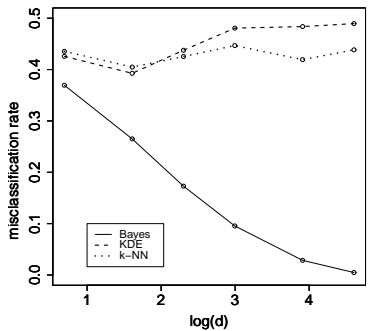}} \hfil
\subfigure[Example {\bf E2}]{\includegraphics[width=5cm]{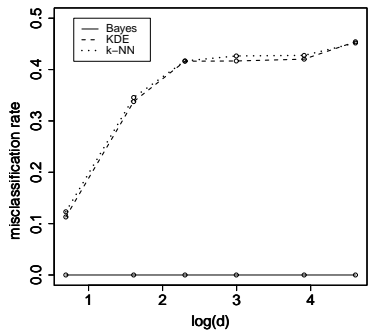}}
\vspace{-.15in}
{\renewcommand{\baselinestretch}{1.25}\caption{\tt Misclassification rates of nonparametric classifiers and the Bayes classifier for $d=2, 5, 10, 20, 50$ and $100$.}}
\vspace{-.2in}
\end{center}
\end{figure}

These two examples clearly show the necessity to develop new classifiers to cope with such situations. Over the last three decades, data depth (see, e.g., \cite{LPS99, ZS00}) has emerged as a powerful tool for data analysis with applications in many areas including supervised and unsupervised classification (see \cite{J04,GC05a,GC05b,HM06,XLY08,DG12,LCL12,LMM12,PVB12}). Spatial depth (also known as the $L_1$ depth) is a popular notion of data depth that was introduced and studied in \cite{VZ00} and \cite{S02}. %It is a real-valued transformation of a multivariate observation that measures the centrality of an observation w.r.t. a multivariate probability distribution.
The {\it spatial depth} (SPD) of an observation $\xvec \in \mathbb{R}^d$ w.r.t. a distribution function $F$ on $\mathbb{R}^d$ is defined as
$\mbox{SPD}(\xvec, F) = 1 - \bigr \|E_F\{ u((\xvec-\Xvec)) \} \bigl \|,$
where $\Xvec \sim F$ and $u(\cdot)$ is the multivariate sign function given by $u(\xvec) = {\|\xvec\|}^{-1}{\xvec}$ if $\xvec \neq {\bf 0}_d \in \mathbb{R}^d$, and $u({\bf 0}_d)={\bf 0}_d$. Henceforth, $\|\cdot\|$ will denote the Euclidean norm. Spatial depth is often computed on the standardized version of the data. In that case, SPD is defined as $$\mbox{SPD}(\xvec, F) = 1 - \bigr \|E_F\{ u(\sigmat^{-1/2}(\xvec-\Xvec)) \} \bigl \|,$$ where $\sigmat$ is a scatter matrix associated with  $F$. If $\sigmat$ has the affine equivariance property, this version of SPD is affine invariant.

Like other depth functions, SPD provides a centre-outward ordering of multivariate data. An observation has higher (respectively, lower) depth if it lies close to (respectively, away from) the centre of the distribution. In other words, given an observation $\xvec$ and a pair of probability distributions $F_1$ and $F_2$, if SPD$(\xvec,F_{1})$ is larger than SPD$(\xvec,F_{2})$, one would expect $\xvec$ to come from $F_1$ instead of $F_2$. Based on this simple idea, the {\it maximum depth classifier} was developed in \cite{GC05b,J04}. For a $J$-class problem involving distributions $F_1, \ldots, F_J$, it classifies an observation $\xvec$ to the $j^*$-th class, where $j^* = \arg \max_{1 \leq j \leq J} \mbox{SPD}(\xvec, F_{j})$.

%{\color{red} We can shift the figures and explanation of maximum SPD to supplmentary material.}

An important property of SPD (see {Lemma 1} in Appendix) is that when the class distribution $F$ is unimodal and spherically symmetric, the class density function turns out to be a monotonically increasing function of SPD. In both examples {\bf E1} and {\bf E2}, the class distributions are spherical. %, and they have the same location but different dispersions and shapes.
Consequently, SPD$(\xvec,F)$ is a function of $\|\xvec\|$ in view of the rotational invariance of SPD$(\xvec,F)$. In Figure 3, we have plotted $\mbox{SPD}(\xvec,F_1)$ and $\mbox{SPD}(\xvec,F_2)$ for different values of $\|\xvec\|$ for examples {\bf E1} and {\bf E2}, where $F_1$ and $F_2$ are the two class distributions and $\xvec \in \mathbb{R}^2$. It is transparent from the plots that the maximum depth classifier based on SPD will fail in both examples. In example {\bf E1}, for all values of $\|\xvec\|$ smaller (respectively, greater) than a constant close to 4, the observations will be classified to the first (respectively, the second) class by the maximum SPD classifier. On the other hand, this classifier will classify all observations to the second class in example {\bf E2}.

In Section 2, we develop a modified classifier based on SPD to overcome this limitation of the maximum depth classifier. Most of the existing modified depth based classifiers are developed mainly for two class problems (see, e.g., \cite{GC05b,DG12,LCL12,PVB12,LMM12}). For classification problems involving $J(>2)$ classes, one usually solves $\binom{J}{2}$ binary classification problems taking one pair of classes at a time and then uses majority votes to make the final classification. Our proposed classification method based on SPD addresses the $J$ class problem directly.

\begin{figure}
\begin{center}
\subfigure[Example {\bf E1}]{\includegraphics[width=5cm]{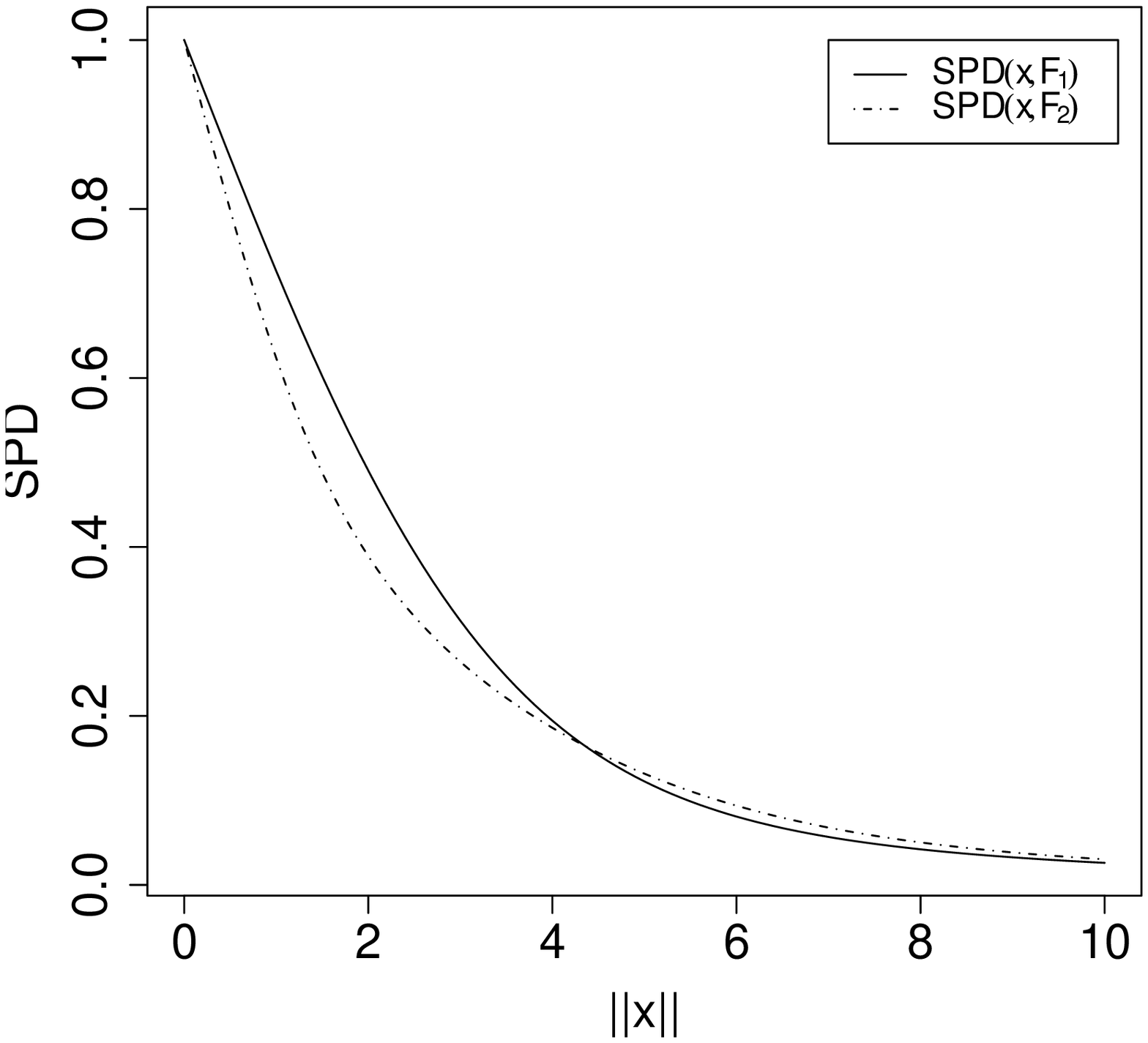}} \hfil
\subfigure[Example {\bf E2}]{\includegraphics[width=5cm]{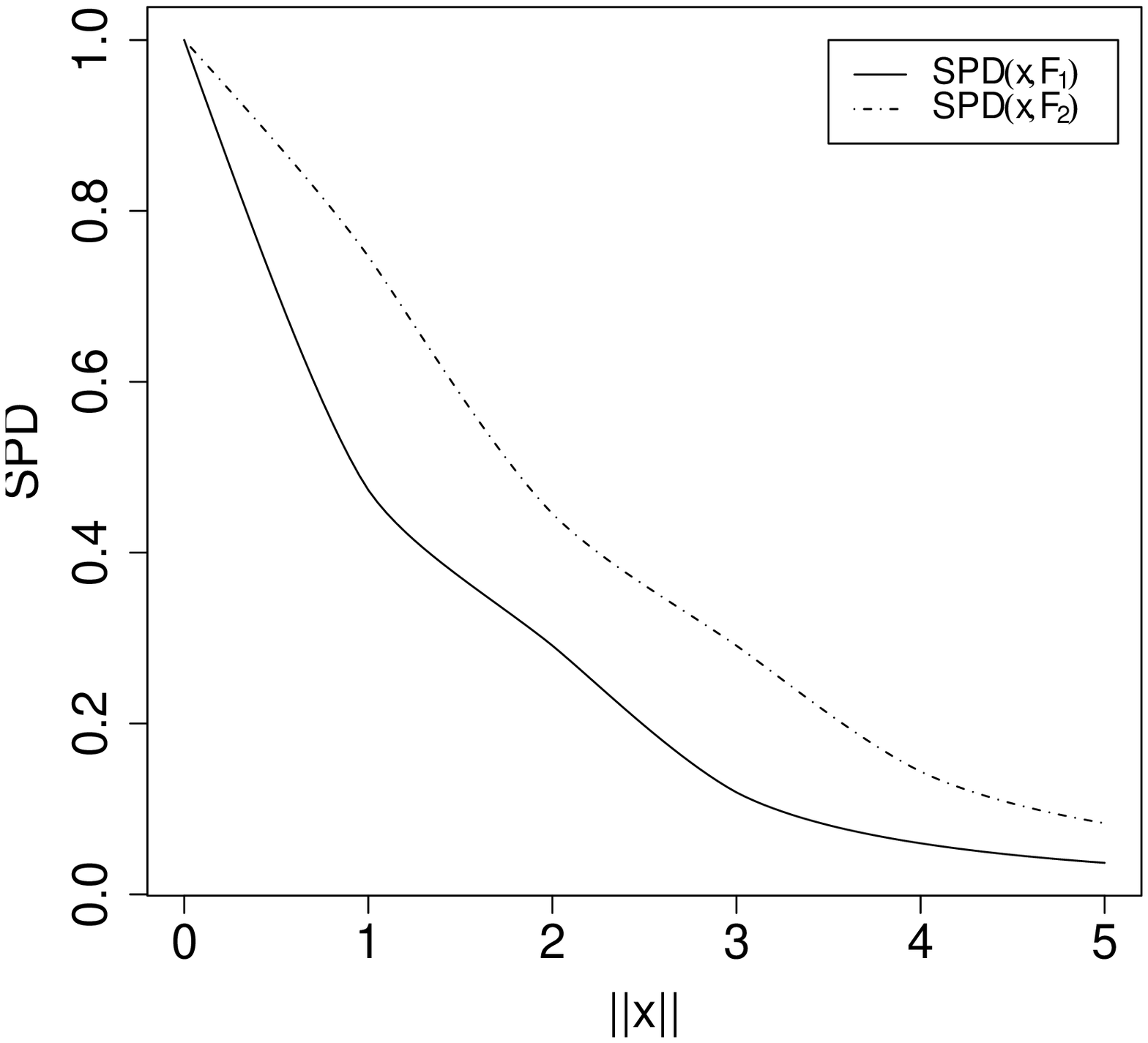}}
\vspace{-.1 in}
\caption{\tt SPD($\xvec, F_1$) and SPD($\xvec, F_2$) for different values of $\|\xvec\|$ when $\xvec \in \mathbb{R}^2$.}
\vspace{-.2in}
\end{center}
\end{figure}

Almost all depth based classifiers proposed in the literature require ellipticity of class distributions to achieve Bayes optimality.
In order to cope with possible multimodal as well as non-elliptic population distributions, we construct a localized version of SPD (henceforth referred to as LSPD) in Section 3. In Section 4, we develop a multiscale classifier based on LSPD. %which turns out to be optimal for a very general class of continuous distributions.
Relevant theoretical results on SPD, LSPD and the resulting classifiers have also been studied in these sections.

An advantage of SPD over other depth functions is its computational simplicity. Classifiers based on SPD and LSPD can be constructed even when the dimension of the data exceeds the sample size. We deal with such high-dimensional low sample size cases in Section 5, and show that both classifiers turn out to be optimal under a fairly general framework. In Sections 6 and 7, some simulated and benchmark data sets are analyzed to establish the usefulness of our classification methods. Section 8 contains a brief summary of the work and some concluding remarks. All proofs and mathematical details are given in the Appendix.

\vspace{-.1in}
\section{Bayes optimality of a classifier based on SPD}%spatial depth in general elliptic populations}

Let us assume that $f_1, \ldots, f_J$ are the density functions of $J$ elliptically symmetric distributions on $\mathbb{R}^d$, where $f_j(\xvec)=|\sigmat_j|^{-1/2}g_j(\|\sigmat_j^{-1/2}(\xvec-\muvec_j)\|)$ for $1 \leq j \leq J$. %, and $\pi_1, \ldots, \pi_J$ are the prior probabilities for the $J$ populations with $\sum_{i=1}^J \pi_i=1$.
 Here $\muvec_j \in \mathbb{R}^d$, $\sigmat_j$ is a $d \times d$ positive definite matrix, and $g_j(\|\tvec\|)$ is a probability density function on $\mathbb{R}^d$ for $1 \leq j \leq J$. For such classification problems involving general elliptic populations with equal or unequal priors, the next theorem establishes the Bayes optimality of a classifier, which is based on $\zvec(\xvec)=(z_1(\xvec), \ldots, z_J(\xvec))^{T}=($SPD$(\xvec, F_1)$, $\ldots$, SPD$(\xvec, F_J))^{T}$, the vector of SPD. In particular, it follows from this theorem that for examples {\bf E1} and {\bf E2} discussed at the beginning of Section 1, the class boundaries (see Figure 1) of the Bayes classifiers are functions of $\zvec(\xvec)=($SPD$(\xvec, F_1)$, SPD$(\xvec, F_2))^{T}$.

\vspace{-.1in}
\newtheorem{theorem}{Theorem}
\begin{theorem}
{\it If the densities of the $J$ competing classes are elliptically symmetric, the posterior probabilities of these classes satisfy the logistic regression model given by
\begin{equation}
p(j|\xvec) = p(j|\zvec(\xvec))= \frac{\exp(\Phi_j(\zvec(\xvec)))}{[1+\sum_{k=1}^{(J-1)}\exp(\Phi_k(\zvec(\xvec)))]} ~\mbox{for}~~ 1 \leq j \leq (J-1)~
\end{equation}
%\vspace{-0.25in}
%for $1 \leq i \leq (J-1)$, and
\begin{equation}
\mbox{and}~~ p(J|\xvec) = p(J|\zvec(\xvec)) = \frac{1}{[1+\sum_{k=1}^{(J-1)}\exp(\Phi_k(\zvec(\xvec)))]}.
\end{equation}
Here $\Phi_j(\zvec(\xvec))=\varphi_{j1}(z_1(\xvec)) + \ldots + \varphi_{jJ}(z_J(\xvec))$, and
$\varphi_{ji}$s are appropriate real-valued functions of real variables.
Consequently, the Bayes rule assigns an observation $\xvec$ to the class $j^*$, where $j^* = arg \; max_{1 \leq j \leq J} \; p(j|\zvec(\xvec))$.}
\end{theorem}
\vspace{-.1in}

{Theorem 1} shows that the Bayes classifier is based on a nonparametric multinomial additive logistic regression model for the posterior probabilities, which is a special case of generalized additive models (GAM) \cite{HT90}. If the prior probabilities of the $J$ classes are equal, and $f_1,\ldots,f_J$ are all elliptic and unimodal differing only in their locations, this Bayes classifier reduces to the maximum SPD classifier \cite{GC05b,J04} (see {Remark 1} after the proof of {Theorem 1} in the Appendix).

For any fixed $i$ and $j$, one can calculate the $J$-dimensional vector $\zvec(\xvec_{ji})$, where $\xvec_{ji}$ is the $i$-th training sample observation in the $j$-th class for $1 \leq i \leq n_j$ and $1 \leq j \leq J$. These $\zvec(\xvec_{ji})$s can be viewed as realizations of the vector of co-variates in a nonparametric multinomial additive logistic regression model, where the response corresponds to the class label that belongs to $\{1, \ldots, J\}$. So, a classifier based on SPD can be constructed by fitting a generalized additive model with the logistic link function.
In practice, when we compute SPD of $\xvec$ from the data $\xvec_{1}, \ldots, \xvec_{n}$ generated from $F$, we use its empirical version as
${\mbox{SPD}}(\xvec, F_n) = 1 - \left \| \frac{1}{n} \sum_{i=1}^{n} u(\xvec - \xvec_i) \right \|$. For the standardized version of the data, it is defined as
$${\mbox{SPD}}(\xvec, F_n) = 1 - \bigg \| \frac{1}{n} \sum_{i=1}^{n} u({\widehat \sigmat}^{-1/2} (\xvec - \xvec_i)) \bigg \|,$$
where ${\widehat \sigmat}$ is an estimate of $\sigmat$, and $F_n$ is the empirical distribution of the data $\xvec_1, \ldots, \xvec_n$. The resulting classifier worked well in examples {\bf E1} and {\bf E2}, and we shall see it in Section 6.
%Note that when ${\widehat \sigmat}$ is an affine equivariant estimate of $\sigmat$, this empirical version of SPD will be affine invariant.

\vspace{-.1in}
\section{Extraction of small scale distributional features by localization of spatial depth}

Under elliptic symmetry, the density function of a class can be expressed as a function of SPD, and hence the SPD contours coincide with the density contours. This is the main mathematical argument used in the proof of {Theorem 1}.  Now, for certain non-elliptic distributions, where the density function cannot be expressed as a function of SPD, such mathematical arguments %used in the earlier section to construct depth based classifiers and to establish their Bayes optimality
are no longer valid. %So, we now check whether the contours of SPD match the density contours for certain non-elliptic distributions.
For instance, consider an equal mixture of N$_d({\bf 0}_d, 0.25{\bf I}_d)$, N$_d(2{\bf 1}_d, 0.25{\bf I}_d)$ and N$_d(4{\bf 1}_d, 0.25{\bf I}_d)$, where ${\bf 1}_d=(1,\ldots,1)^{T}$. We have plotted its SPD contours in Figure 4 when $d=2$. For this trimodal distribution, the SPD contours fail to match the density contours. As a second example, we consider a $d$-dimensional distribution with independent components, where the $i$-th component is exponential with the scale parameter $d/(d-i+1)$ for $1 \leq i \leq d$. We have plotted its SPD contours in Figure 5 when $d=2$. Even in this example, the SPD contours differ significantly from the density contours. To cope with this issue, we suggest a {\it localization} of SPD (see the third contour plots $(c)$ in Figures 4 and 5). As we shall see later, this localized SPD relates to the underlying density function, and the resulting classifier turns out to be the Bayes classifier (in a limiting sense) in a general nonparametric setup with arbitrary
class densities.

\begin{figure*}[!t]
\begin{center}
{
\subfigure[Density]{\includegraphics[angle=-90,width=5cm]{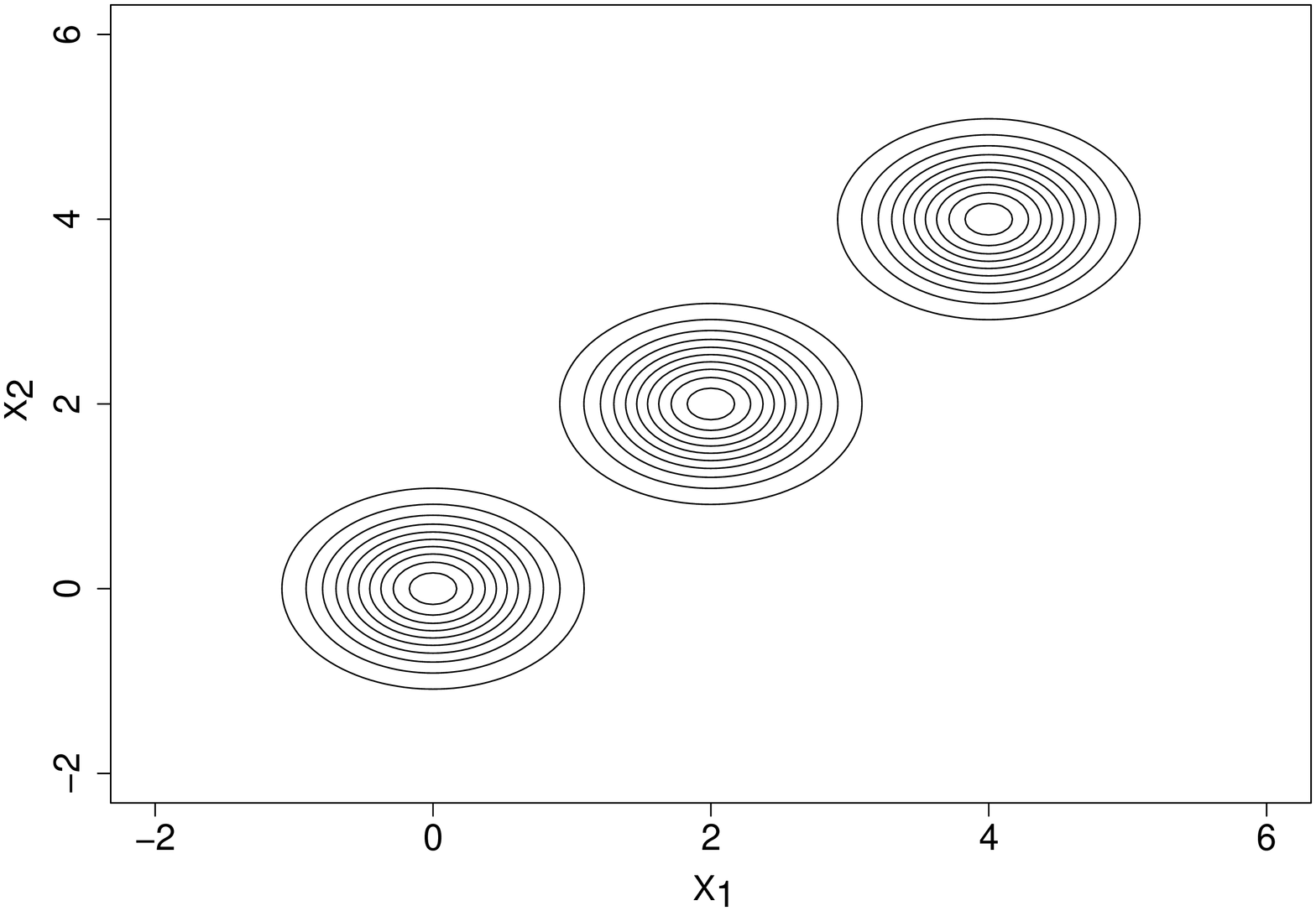} \hspace{-.1in}} \hfil
\subfigure[SPD]{\includegraphics[angle=-90,width=5cm]{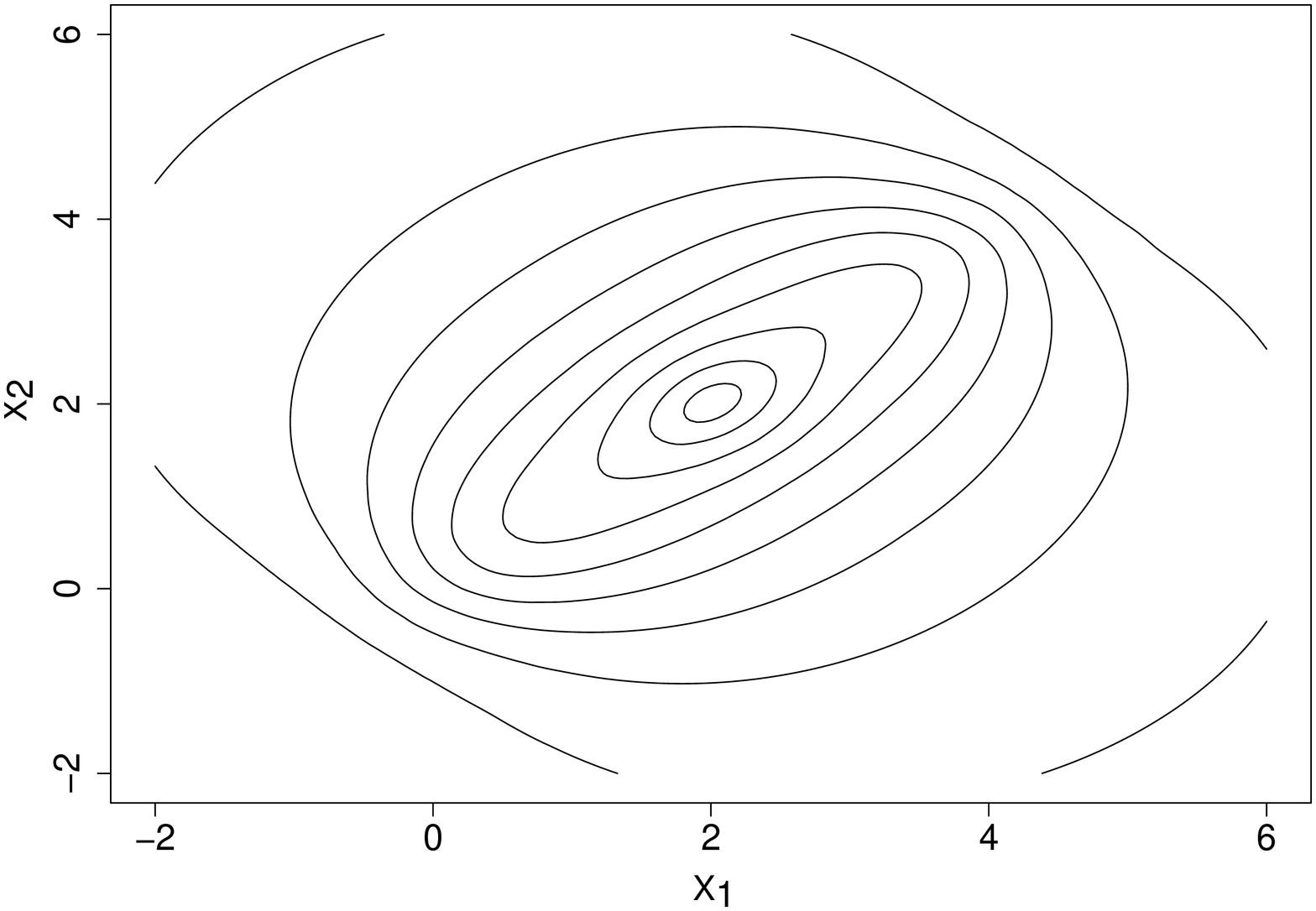} \hspace{-.1in}} \hfil
\subfigure[LSPD$_{h=.4}$]{\includegraphics[angle=-90,width=5cm]{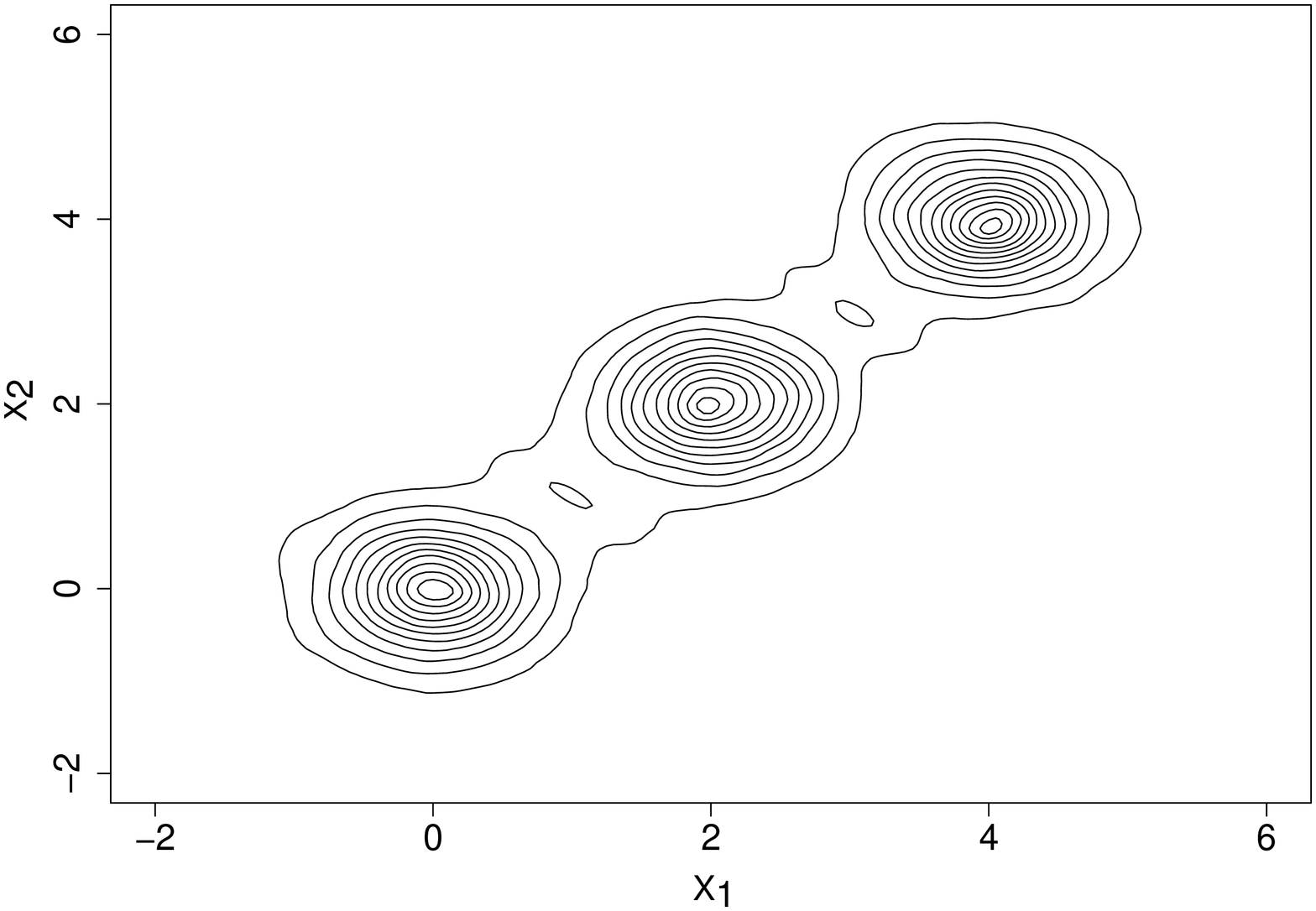}}
}
\vspace{-0.1in}
{\renewcommand{\baselinestretch}{1.25}\caption{\tt Contours of density, SPD and LSPD$_h$ (with $h=.4$) functions for a symmetric, trimodal density function.}}
\end{center}
%\end{figure*}
%
%\begin{figure*}
\vspace{-0.25in}
\begin{center}
{
\subfigure[Density]{\includegraphics[angle=-90,width=5cm]{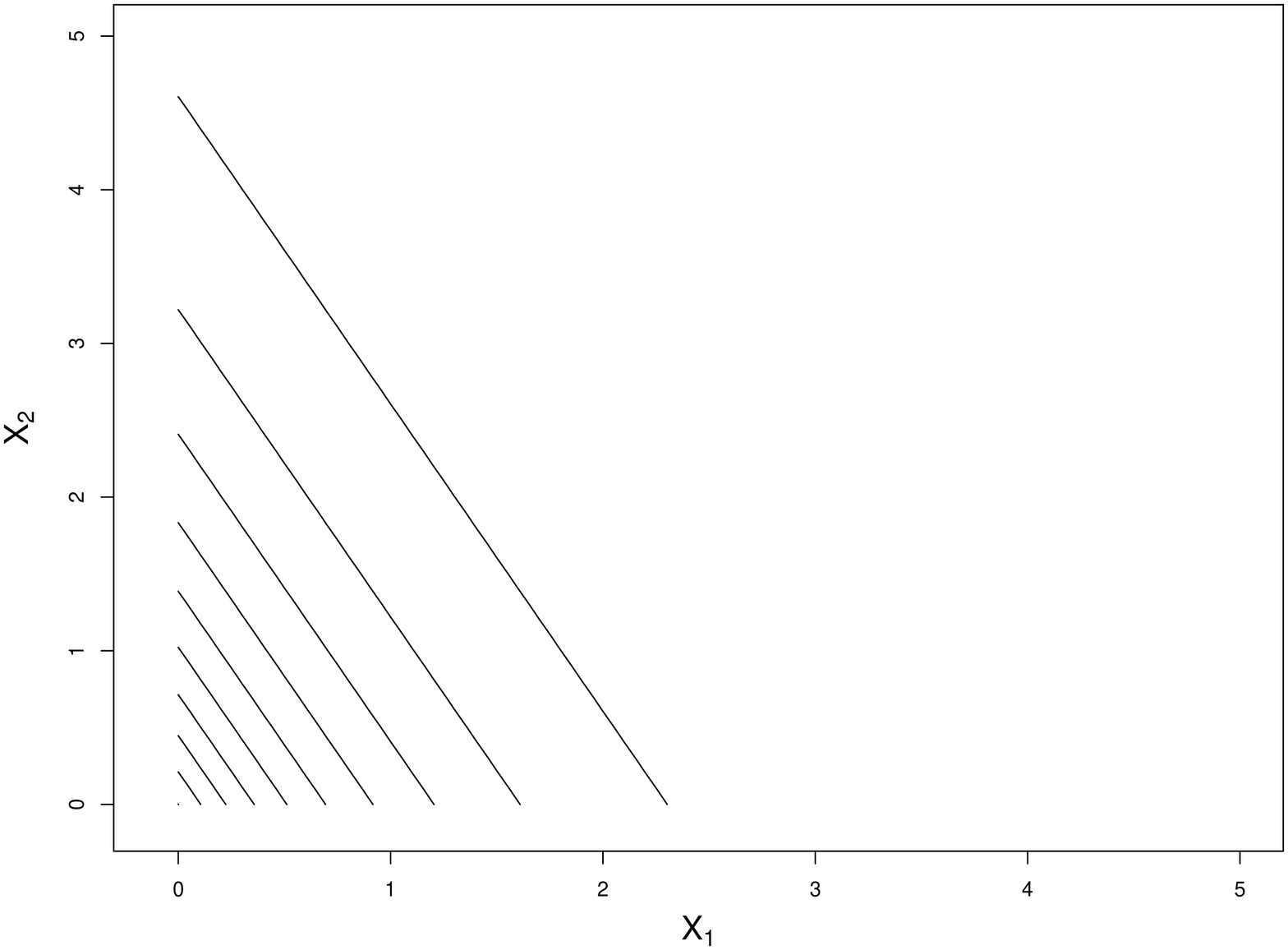} \hspace{-.1in}} \hfil
\subfigure[SPD]{\includegraphics[angle=-90,width=5cm]{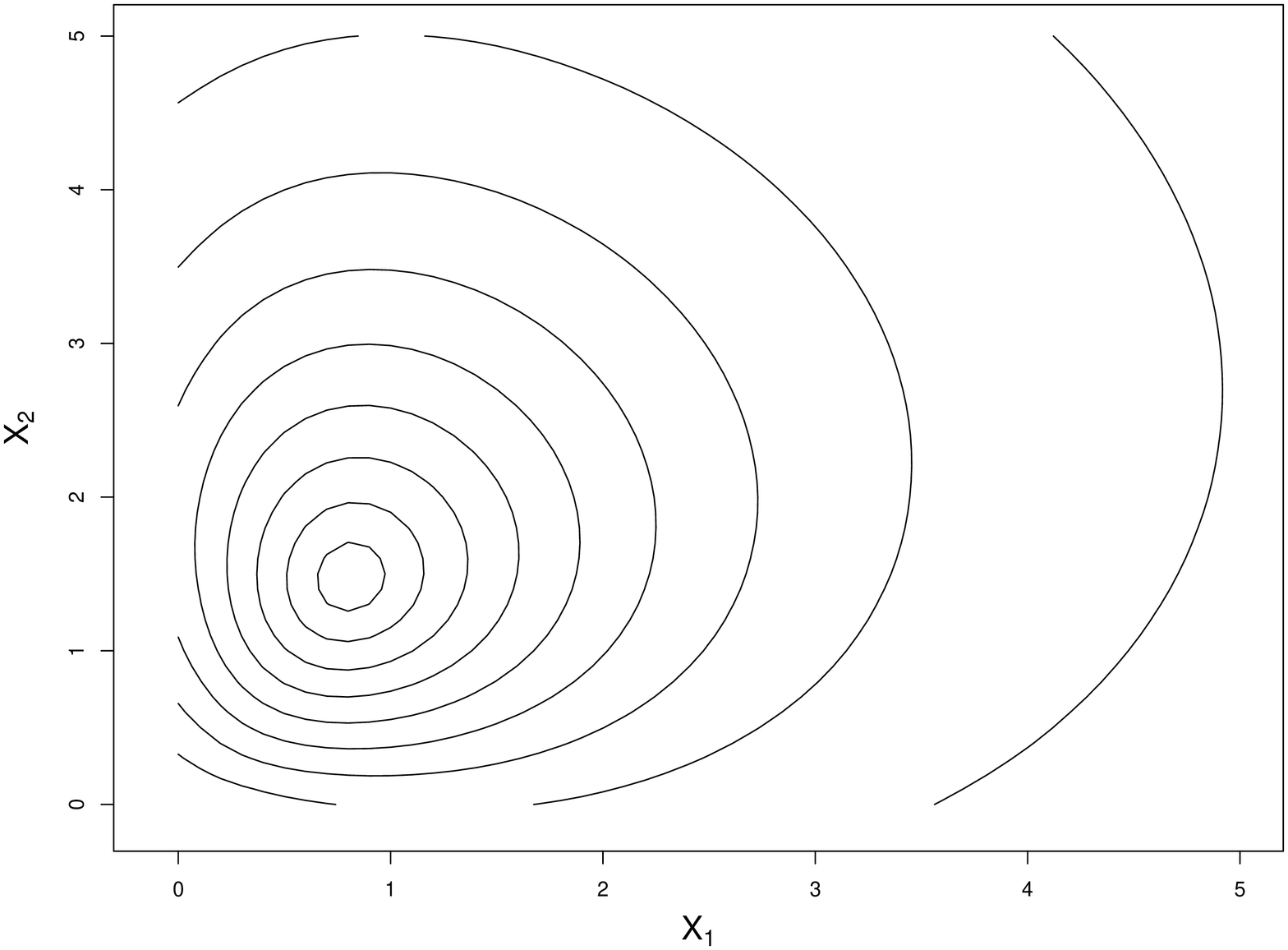} \hspace{-.1in}} \hfil
\subfigure[LSPD$_{h=.25}$]{\includegraphics[angle=-90,width=5cm]{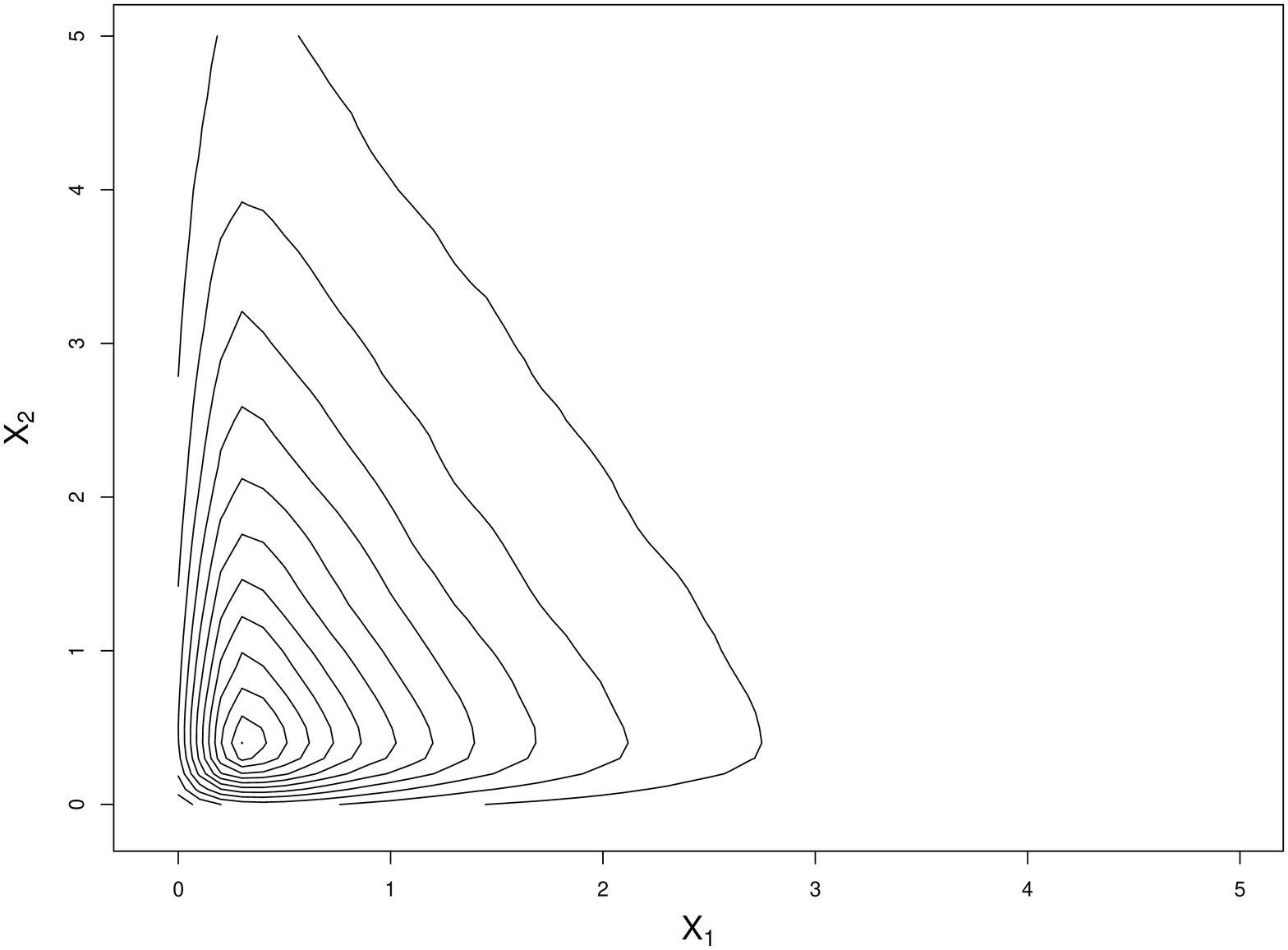}}
}
\vspace{-0.1in}
{\renewcommand{\baselinestretch}{1.25}\caption{\tt Contours of density, SPD and LSPD$_h$ (with $h=.25$) functions for the density function $f(x_1,x_2) = .5 \exp \{-(x_1+.5x_2)\}I\{x_1>0,x_2>0\}$.}}
\end{center}
\vspace{-.1in}
\end{figure*}

Note that $\mbox{SPD}(\xvec, F) = 1 - \| E_F \{ u(\xvec - \Xvec) \} \|$ is constructed by assigning the same weight to each unit vector $u(\xvec-\Xvec)$ ignoring the significance of distance between $\xvec$ and $\Xvec$. By introducing a weight function, which depends on this distance, one can extract important features related to the local geometry of the data. To capture these local features, we introduce a kernel function $K(\cdot)$ as a weight and define
$$\Gamma_h(\xvec,F) = E_F[{K}_h(\tvec)] - \| E_F[K_h(\tvec) u(\tvec)] \|,$$
where $\tvec = (\xvec - \Xvec)$ and $K_h(\tvec) = h^{-d} K(\tvec/h)$. Here $K$ is chosen to be a bounded continuous density function on $\mathbb{R}^d$ such that $K(\tvec)$ is a decreasing function of $\|\tvec\|$ and $K(\tvec) \rightarrow 0$ as $\|\tvec\| \rightarrow \infty$. The Gaussian kernel $K(\tvec) = (\sqrt{2 \pi})^{-d} \exp\{-\|\tvec\|^2/2\}$ is a possible choice. %In $\| E_F[K_h(\tvec) u(\tvec)] \|$, we use a weight function which depends on the Mahalanobis distance $\|{\sigmat}^{-1/2}(\xvec - \Xvec)\|$ of $\Xvec$ from $\xvec$; and $E_F[{K}_h(\tvec)]$ can be viewed as the average weight, where the averaging is done using the probability distribution $F$.
It is desirable that the localized version of SPD approximates the class density or a monotone function of it for small values of $h$. This will ensure that the class densities and hence, the class posterior probabilities become functions of the local depth as $h \rightarrow 0$. On the other hand, one should expect that as $h \rightarrow \infty$, the localized version of SPD should tend to ${\mbox{SPD}}$ or a monotone function of it. However, ${\Gamma}_h(\xvec,F) \rightarrow 0$ as $h \rightarrow \infty$. So, we re-scale $\Gamma_h(\xvec,F)$ by an appropriate factor to define the {\it localized spatial depth} (LSPD) function as follows:
\vspace{-.01in}
\begin{equation}
{\mbox{LSPD}}_h(\xvec,F) = \left \{
\begin{array}{ll}
\Gamma_h(\xvec,F) & \mbox{if}~ h \leq 1, \\
h^{d} \Gamma_h(\xvec,F) & \mbox{if}~ h > 1.
\end{array}
\right.
\vspace{-.01in}
\end{equation}
%Instead of $\tvec=(\xvec-\Xvec)$,
Using $\tvec=\sigmat^{-1/2}(\xvec-\Xvec)$ in the definition of ${\Gamma}_h(\xvec,F)$, one gets LSPD on standardized data, which is affine invariant if $\sigmat$ is affine equivariant. $\mbox{LSPD}_h$ defined this way is a continuous function of $h$, and
$\zvec_h(\xvec) = ($LSPD$_h(\xvec, F_1)$, $\ldots$, LSPD$_h(\xvec, F_J))^{T}$ has the desired behavior as shown in {Theorem 2}.

\vspace{-.1in}
\noindent
\begin{theorem}
{\it Consider a kernel function $K(\tvec)$ that satisfy $\int_{\mathbb{R}^d} \|\tvec\| K(\tvec) d\tvec < \infty$. If $f_1, \ldots, f_J$ are continuous density functions with bounded first derivatives, and the scatter matrix $\sigmat_{j}$ corresponding to $f_j(\xvec)$ exists for all $1 \leq j \leq J$, then

\noindent
$(a) ~\zvec_h(\xvec) \rightarrow (|\sigmat_{1}|^{1/2}f_1(\xvec), \ldots, |\sigmat_{J}|^{1/2}f_J(\xvec))^T~\mbox{as}~h \rightarrow 0$, and

\noindent
$(b) ~\zvec_h(\xvec) \rightarrow (K({\bf 0})\mbox{SPD}(\xvec, F_1), \ldots, K({\bf 0})\mbox{SPD}(\xvec, F_J))^T ~\mbox{as}~ h \rightarrow \infty.$
}
\end{theorem}
\vspace{-.1in}

Now, we construct a classifier by plugging in LSPD$_h$ instead of SPD in the GAM discussed in Section 2. %So, for $\zvec_h(\xvec) = ($LSPD$_h(\xvec, F_1)$, $\ldots$, LSPD$_h(\xvec, F_J))^{T}$, we
So, we consider the following model for the posterior probabilities
\begin{equation}
p(j|\zvec_h(\xvec)) = \frac{\exp(\Phi_j(\zvec_h(\xvec)))}{[1+\sum_{k=1}^{(J-1)}\exp(\Phi_k(\zvec_h(\xvec)))]}, ~\mbox{for}~1 \le j < (J-1),
\end{equation}
\begin{equation}
\mbox{and}~~p(J|\zvec_h(\xvec)) = \frac{1}{[1+\sum_{k=1}^{(J-1)}\exp(\Phi_k(\zvec_h(\xvec)))]}.
\end{equation}

The main implication of part $(a)$ of {Theorem 2} is that the classifier constructed using GAM and $\zvec_h(\xvec)$ as the covariate tends to the Bayes classifier in a general nonparametric setup as $h \rightarrow 0$. On the other hand, part $(b)$ of {Theorem 2} implies that for elliptic class distributions, the same classifier tends to the Bayes classifier when $h \rightarrow \infty$. When we fit GAM, the functions $\Phi_j$s are estimated nonparametrically. Flexibility of such nonparametric estimates automatically takes care of the constants $|\sigmat_{j}|^{1/2}$ for $1 \leq j \leq J$ and $K({\bf 0})$ in the expressions of the limiting values of $\zvec_h(\xvec)$ in parts $(a)$ and $(b)$ of {Theorem 2}, respectively.

The empirical version of $\Gamma_h(\xvec,F)$, denoted by ${\Gamma_h}(\xvec, F_n)$, is defined as
$${\Gamma_h}(\xvec, F_n) = \frac{1}{n} \sum_{i=1}^{n} K_h(\tvec_i) - \biggl \| \frac{1}{n} \sum_{i=1}^{n} K_h(\tvec_i) u{(\tvec_i)} \biggr \|,$$
where $\tvec_i =(\xvec-\xvec_i)$ (or, ${\widehat \sigmat}^{-1/2}(\xvec-\xvec_i)$ if we use standardized version of the data) for $1 \leq i \leq n$. Then ${\mbox{LSPD}}_h(\xvec, F_n)$ is defined using $(3)$ with ${\Gamma_h(\xvec,F)}$ replaced by ${\Gamma_h}(\xvec, F_n)$. Theorem 3 below shows the almost sure uniform convergence of ${\mbox{LSPD}}_h(\xvec, F_n)$ to its population counterpart ${\mbox{LSPD}}_h(\xvec, F)$. Similar convergence result for the empirical version of SPD has been proved in the literature (see, e.g., \cite{G03}).

\vspace{-0.1in}
\begin{theorem}
{\it Suppose that the density function $f$ and the kernel $K$ are bounded, %, and has bounded first derivatives.
and $K$ has bounded first derivatives. %Also assume that $\int \|\tvec\|K(\tvec) d\tvec <\infty$.
%, and $\int \|\tvec\|K(\tvec)d\tvec$ is finite.
Then, for any fixed $h>0$,
$\sup_{\xvec} |\mbox{LSPD}_{h}(\xvec,$ $F_{n}) - \mbox{LSPD}_{h}(\xvec,F)| \stackrel{a.s.}{\rightarrow} 0~\mbox{as}~n \rightarrow \infty$.}
\vspace{-0.1in}
\end{theorem}
%\begin{theorem}
%Under suitable conditions,

%\noindent
%$(a) ~({\mbox{LSPD}}_h(\xvec,F_{1n_1}),\ldots,{\mbox{LSPD}}_h(\xvec,F_{Jn_J}) \stackrel{P}{\rightarrow} (f_1(\xvec),\ldots,f_J(\xvec))~\mbox{as}~h \to 0$, and

%\noindent
%$(b) ~({\mbox{LSPD}}_h(\xvec,F_{1n_1}),\ldots,{\mbox{LSPD}}_h(\xvec,F_{Jn_J}) \stackrel{P}{\rightarrow} K({\bf %0})(SPD(\xvec,F_1),\ldots,SPD(\xvec,F_J))~\mbox{as}~h \to \infty$.
%\end{theorem}

%TO the best of our knowledge, we are not aware of any consistency results for procedures based on GAM. Such a result would prove the consistency of the overall classification procedure. However, from the simulations, empirically it is evident that the performance of our method is quite good w.r.t. the Bayes risk.

%We can extend this proof to all $J$ classes by applying Theorem 3 to each component.
From the proof of Theorem 3, it is easy to check that this almost sure uniform convergence also holds when $h \rightarrow \infty$. Under additional moment conditions on $f$ and $K$, this holds for the $h \rightarrow 0$ case as well if $nh^{2d}/\log n \rightarrow \infty$ as $n \rightarrow \infty$ (see Remarks 2 and 3 after the proof of Theorem 3 in the Appendix). So, the result stated in parts (a) and (b) of Theorem 2 continue to hold for the empirical version of LSPD under appropriate assumptions.

Localization and kernelization of different notions of data depth have been considered in the literature
%earlier by several authors including
\cite{CDPB09, AR10, LL11, HWWLH11, PVB13}. The fact that LSPD$_h$ tends to a constant multiple of the probability density function as $h \rightarrow 0$ is a crucial requirement for limiting Bayes optimality of classifiers based on this localized depth function. In \cite{AR10}, the authors proposed localized versions of simplicial depth and half-space depth, but the relationship between the local depth and the probability density function has been established only in the univariate case. A depth function based on inter-point distances has been developed in \cite{LL11} to capture multimodality in a data set. %, but there is no specific result regarding its relation with the corresponding density function.
%that can ensure that this depth function can capture the density for appropriate choices of the localization parameter.
Chen {\it et al.} \cite{CDPB09} defined kernelized spatial depth in a reproducing kernel Hilbert space. In \cite{HWWLH11}, the authors considered a generalized notion of Mahalanobis depth in reproducing kernel Hilbert spaces.
%and they empirically demonstrated that it %the ``kernelized spatial depth"
%can capture interesting and complex features of the data in many situations.
%But, there is no specific result %regarding their relations with the corresponding density function.
However, there is no result connecting them to the probability density function. Infact, the kernelized spatial depth function becomes degenerate at the value $(1-1/{\sqrt{2}})$ as the tuning parameter goes to zero. Consequently, it becomes non-informative for small values of the tuning parameter.
%In \cite{HWWLH11}, the authors considered a generalized notion of Mahalanobis depth in reproducing kernel Hilbert spaces. %They gave numerical demonstrations for the influence of the localization parameter on various non-convex data sets.
% ,
%But, again  there is no result connecting this depth to the density function.
It will be appropriate to note here that none of the preceding authors used their proposed depth functions for constructing classifiers.
%in supervised classification problems.
Recently, in \cite{PVB12, PVB13}, the authors proposed a notion of local depth and used it for supervised classification. But, their proposed version of local depth does not relate to the underlying density function either.

%{\color{red} We can make this review shorter.}

\vspace{-.1in}
\section{Multiscale classification %by aggregating posteriors
based on LSPD}

When the class distributions are elliptic, part $(b)$ of {Theorem 2} implies that LSPD$_h$ with appropriately large choices of $h$ lead to good classifiers. These large values may not be appropriate for non-elliptic class distributions, but part $(a)$ of {Theorem 2} implies that LSPD$_h$
with appropriately small choices of $h$ lead to good classifiers for general nonparametric models for class densities. However, for small values of $h$, the empirical version of LSPD$_h$ behaves like a nonparametric density estimate, and it suffers from the curse of dimensionality. So, the resulting classifier may have its statistical limitations for high-dimensional data.

\begin{figure}[!b]
\vspace{-.25in}
\begin{center}
\hspace{-.1in}
\includegraphics[height=6cm,width=7cm]{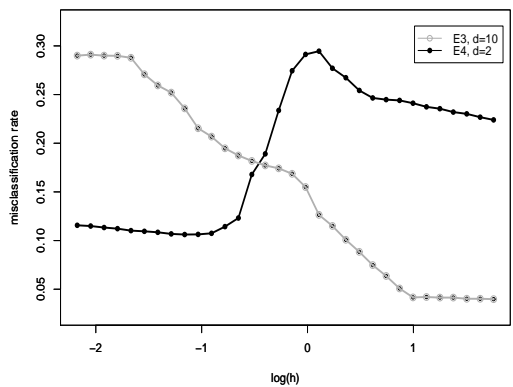}
\vspace{-.225in}
\caption{\tt Misclassification rates in examples {\bf E3} and {\bf E4} for the classifier based on ${\mbox{LSPD}}_h$ for different values of $h$.}
\vspace{-.2in}
\end{center}
\end{figure}

We now consider two examples to demonstrate the above points. The first example (we call it {\bf E3}) involves two multivariate normal distributions %differing in location as well as scale. The first one is
N$_{d}({\bf 0}_d, {\bf I}_{d})$ and %the second one is
N$_{d}({\bf 1}_d, 4{\bf I}_{d})$.
%, where ${\bf 1}_d=(1,\ldots,1)^{T} \in \mathbb{R}^{d}$, ${\bf 0}_d=(0,\ldots,0)^{T} \in \mathbb{R}^{d}$ and ${\bf I}_{d}$ is the $d \times d$ identity matrix as before.
In the second example (we call it {\bf E4}), both distributions are {trimodal}. The first class has the same density as in Figure 4 (i.e., an equal mixture of N$_d({\bf 0}_d, 0.25{\bf I}_d)$, N$_d(2{\bf 1}_d, 0.25{\bf I}_d)$ and N$_d(4{\bf 1}_d, 0.25{\bf I}_d)$), while the second class is an equal mixture of N$_d({\bf 1}_d, 0.25{\bf I}_d)$, N$_d(3{\bf 1}_d, 0.25{\bf I}_d)$ and N$_d(5{\bf 1}_d, 0.25{\bf I}_d)$. We consider the case $d=10$ for {\bf E3} and $d=2$ for {\bf E4}. For each of these two examples, we generated a training sample of size 100 from each class. The misclassification rate for the classifier based on ${\mbox{LSPD}}_h$ was computed based on a test sample of size 500 (250 from each class). This procedure was repeated 100 times to calculate the average misclassification rate for different values of $h$. Figure 6 shows that the large (respectively, small) values of $h$ yielded low misclassification rates in {\bf E3} (respectively, {\bf E4}). For small values of $h$, empirical LSPD$_h$ behaved like a nonparametric density estimate that suffered from the curse of dimensionality in {\bf E4}. Consequently, its performance deterioratesd. But, for large $h$, the underlying elliptic structure was captured well by the proposed classifier. This provides a strong motivation for using a multi-scale approach in constructing the final classifier so that one can harness the strength of different classifiers corresponding to different levels of localization of SPD. One would expect that when aggregated, classifiers corresponding to different values of $h$ will lead to improved misclassification rates. Usefulness of the multi-scale approach in combining different classifiers has been discussed in the classification literature by several authors including \cite{DZ04,GCM05,GCS06,KHDM98}.

%{\color{red} We need to talk about which version(s) of LSPD to put here}

A popular way of aggregation is to consider the weighted average of the estimated posterior probabilities computed for different values of $h$. There are various proposals for the choice of the weight function in the literature. Following \cite{GCM05}, one can compute
${\widehat {\Delta}}_{h}$, the leave-one-out estimate of the misclassification rate of the classifier based on LSPD$_h$ and use
$$\centering{ W(h) \propto \exp \left[-\frac{1}{2} \frac{({\widehat \Delta}_{h} - {\widehat \Delta}_{0})^2} {{\widehat \Delta}_{0} (1-{\widehat \Delta}_{0})/n} \right]}$$ as the weight function, where ${\widehat \Delta}_0 = \disp \min_{h} {\widehat \Delta}_{h}$. %, and $C$ is a normalizing constant such that $\sum_h W(h) = 1$.
The exponential function helps to appropriately weighing up (respectively, down) the promising (respectively, poor) classifier resulting from different choices of the smoothing parameter $h$. However, $\int W(h) dh$ or $\int p(j|z_h(\xvec)) W(h) dh$ may not be finite for some choices of $j \in \{1,2,\ldots,J\}$. So, here we use a slightly modified weight function $W^{*}(h)=W(h)g(h)$, where $g$ is a univariate Cauchy density with a
large scale parameter and support restricted to have positive values only.
 %Following \cite{GCM05}, we can choose a weight function $W(h)$, which involves estimates of the misclassification rates for different values of the localization parameter $h$.
Our final classifier, which we call the LSPD classifier, assigns an observation $\xvec$ to the $j^*$-th class, where
\vspace{-.05in}
$$j^*=\arg \max_{1 \leq j \leq J}  \int_{h>0} W^{*}(h)~p(j|\zvec_h(\xvec))dh= \arg \max_{1 \leq j \leq J}\int_{h>0} W(h)g(h)~p(j|\zvec_h(\xvec))dh.
\vspace{-.05in}$$
Here $p(j|\zvec_h(\xvec))$ is as in equations (4) and (5) in Section 3. In practice, we first generate $M$ independent observations $h_1,h_2,\ldots,h_M$ from $g$. For any given $j$ and $\xvec$, we approximate $\int_{h>0} W(h)g(h)~p(j|\zvec_h(\xvec))dh$ by $\sum_{i=1}^{M} W(h_i)~p(j|\zvec_{h_i}(\xvec))/M$. The use of the Cauchy distribution with a large scale parameter (we use 100 in this article) helps us to generate small as well as large values of $h$. This is desirable in view of Theorem 2.
%Throughout this article, for our numerical work, we use the scale parameter $100$.

%In our numerical work, following \cite{CGO09}, we set $L$ to be the lower fifth percentile of the pairwise distances of the standardized observations over all $J$ classes and $U$ to be the smallest value of $h$ for which $\max_{i,j} \| \hat \zvec_h(\xvec_{ij}) - K({\bf 0}) \hat \zvec(\xvec_{ij}) \| < 0.01$.  Here $\hat{\zvec}(\xvec_{ij})=({\mbox{SPD}}(\xvec_{ij}, F_{n_1}), \ldots, {\mbox{SPD}}(\xvec_{ij}, F_{n_J}))$ and $\hat{\zvec}_h(\xvec_{ij})=({\mbox{LSPD}}_h(\xvec_{ij}, F_{n_1}), \ldots, {\mbox{LSPD}}_h(\xvec_{ij}, F_{n_J}))$ are defined as before.

\vspace{-.16in}
\section{Classification of high-dimensional data}

A serious practical limitation of many existing depth based classifiers is their computational complexity in high dimensions, and this makes such classifiers impossible to use even for moderately large dimensional data. Besides, depth functions that are based on random simplices formed by the data points (see \cite{LPS99,ZS00}), cannot be defined in a meaningful way if dimension of the data exceeds the sample size. Projection depth and Tukey's half-space depth (see, e.g., \cite{ZS00}) both become degenerate at zero for such high-dimensional data. Classification of high-dimensional data presents a substantial challenge to many nonparametric classification tools as well. We have seen in examples {\bf E1} and {\bf E2} (see Figure 2) that nonparametric classifiers like those based on $k$-NN and KDE can yield poor performance when data dimension is large. Some limitations of support vector machines for classification of high-dimensional data have also been noted in \cite{DWD}.

One of our primary motivations behind using SPD is its computational tractability, especially when the dimension is large. We now investigate the behavior of classifiers based on SPD and LSPD for such high-dimensional data. For this investigation, we assume that the observations are all standardized by a common positive definite matrix $\sigmat$ for all $J$ classes, and the following conditions are stated for those standardized random vectors, which are written as $\Xvec$s for notational convenience.

%\vspace{.05in}
(C1) Consider a random vector $\Xvec_1 = (X_1^{(1)}, \ldots, X_1^{(d)})^{T}$ $\sim F_j$. Assume that $a_j = \lim_{d \rightarrow \infty} d^{-1}$ $\sum_{k=1}^{d} E(X_{1}^{(k)})^2$ exists for $1 \leq j \leq J$, and
$d^{-1} \sum_{k=1}^{d} (X_{1}^{(k)})^2 \stackrel{a.s.}{\rightarrow} a_{j}~\mbox{as}~d \rightarrow \infty.
%\vspace{-.05in}
$

%\vspace{-0.1in}
(C2) Consider two independent random vectors $\Xvec_1= (X_1^{(1)}, \ldots, X_1^{(d)})^{T} \sim F_j$ and $\Xvec_2= (X_2^{(1)}, \ldots, X_2^{(d)})^{T} \sim F_i$. Assume that $b_{ji} = \lim_{d \rightarrow \infty}$ $d^{-1} \sum_{k=1}^{d} E(X_{1}^{(k)}X_{2}^{(k)})$ exists,
and $d^{-1} \sum_{k=1}^{d} X_{1}^{(k)} X_{2}^{(k)} \stackrel{a.s.}{\rightarrow} b_{ji}~\mbox{as}~d \rightarrow \infty$ for all $1 \leq j,i \leq J$.

%\vspace{-0.1in}
It is not difficult to verify that for $\Xvec_1 \sim F_j$ ($1 \leq j \leq J$), if we assume that the sequence of variables $\{X_{1}^{(k)} - E(X_{1}^{(k)}) : k=1, 2, \ldots \}$ centered at their means are independent with uniformly bounded eighth moments (see Theorem 1 (2) in \cite{JM09}, p. 4110), or if we assume that they are $m$-dependent random variables with some appropriate conditions (see Theorem 2 in \cite{dJ95}, p. 350), then the almost sure convergence in (C1) as well as (C2) holds. As a matter of fact, the almost sure convergence stated in (C1) and (C2) holds if we assume that for all $1 \leq j,i \leq J$, the sequences $\{(X_{1}^{(k)})^2 - E(X_{1}^{(k)})^2 : k=1, 2, \ldots \}$ and $\{X_{1}^{(k)} X_{2}^{(k)} - E(X_{1}^{(k)}X_{2}^{(k)}) : k=1, 2, \ldots \}$, where $\Xvec_1 \sim F_j$ and $\Xvec_2 \sim F_i$, are {\it mixingales} satisfying some appropriate conditions (see, e.g., Theorem 2 in \cite{dJ95},\;p. 350). Define $\sigma_j^2 = a_j-b_{jj}$ and $\nu_{ji} = b_{jj}-2b_{ji}+b_{ii}$. For the random
vector $\Xvec_1 \sim F_j$, $\sigma_j^2$ is the limit of $d^{-1} \sum_{k=1}^{d} V(X_{1}^{(k)})$ as $d \rightarrow \infty$, where $V(Z)$ denotes the variance of a random variable $Z$. If we consider a second independent random vector $\Xvec_2 \sim F_i$ with $i \neq j$, then $\nu_{ji}$ is the limit of $d^{-1} \sum_{k=1}^{d} \{E(X_{1}^{(k)})-E(X_{2}^{(k)})\}^2$ as $d \rightarrow \infty$. In \cite{HMN05}, the authors assumed a similar set of conditions to study the performance of the classifier based on support vector machines (SVM) with a linear kernel and the $k$-NN classifier with $k=1$ as the data dimension grows to infinity. Similar conditions on observation vectors were also considered in \cite{JM09} to study the consistency of principal components of the sample dispersion matrix for high-dimensional data. Under (C1) and (C2), the following theorem describes the behavior of $\zvec(\xvec)$ and $\zvec_h(\xvec)$ as $d$ grows to infinity.

\vspace{-.1 in}
\noindent
\begin{theorem}
{\it Suppose that the conditions (C1)-(C2) hold, and $\Xvec \sim F_j$ $(1 \le j \le J)$.

\noindent
(a) $\zvec(\Xvec) \stackrel{a.s.}{\rightarrow} (c_{j1}, \ldots, c_{jJ})^{T}=\cvec_j~\mbox{as}~d \rightarrow \infty$,
where $c_{jj} = 1-\sqrt{\frac{1}{2}}$ and $c_{ji} = 1-\sqrt{\frac{\sigma_j^2+\nu_{ji}}{\sigma_j^2+\sigma_i^2+\nu_{ji}}}$ for $1 \leq j \neq i \leq J$.

\noindent
(b) Assume that $h \rightarrow \infty$ and $d \rightarrow \infty$ in such a way that ${\sqrt{d}}/{h} \rightarrow 0~\mbox{or}~A (> 0)$. Then,
$\zvec_h(\Xvec) \stackrel{a.s.}{\rightarrow} g(0)\cvec_j~\mbox{or}~\cvec_j^{\prime}=(g(e_{j1}A) c_{j1}, \ldots, g(e_{jJ}A) c_{jJ})^{T}$
depending on whether ${\sqrt{d}}/{h} \rightarrow 0~\mbox{or}~A$, respectively.
Here $K(\tvec) = g(\|\tvec\|)$, $e_{jj}=\sqrt{2}\sigma_j$ and $e_{ji}=\sqrt{\sigma_j^2+\sigma_i^2+\nu_{ji}}$ for $ j \neq i$.

\noindent
(c) Assume that $h>1$, and ${\sqrt{d}}/{h} \rightarrow \infty$ as $d \rightarrow \infty$.
Then, $\zvec_h(\Xvec) \stackrel{a.s.}{\rightarrow} {\bf 0}_J$.
}
\end{theorem}
\vspace{-.1 in}

The $\cvec_j$s as well as the $\cvec_j^{\prime}$s in the statement of Theorem 4 are distinct for all $1 \leq j \leq J$ whenever either $\sigma_j^2 \neq \sigma_i^2$ or $\nu_{ji} \neq 0$ for all $1 \leq j \neq i \leq J$ (see Lemma 2 in Appendix). In such a case, part $(a)$ of {Theorem 4} implies that for large $d$, $\zvec(\xvec)$  has good discriminatory power, and our classifier based on SPD can discriminate well among the $J$ populations. Further, it follows from part $(b)$ that when both $d$ and $h$ grow to infinity in such a way that $\sqrt{d}/h \rightarrow 0$ or a positive constant, $\zvec_h(\xvec)$ has good discriminatory power as well, and our classifier based on LSPD$_h$ can yield  low misclassification probability. %So, in order to have low misclassification rates, one should aggregate classifiers based on LSPD$_h$ over different choices of $h$ to arrive at a final decision.
However, part $(c)$ shows that if $\sqrt{d}$ grows at a rate faster than $h$, ${\zvec}_h(\xvec)$  becomes non-informative. Consequently, the classifier based on LSPD$_h$ lead to high misclassification probability in this case.

% \vspace{-.1in}
\section{Analysis of simulated data sets}

\noindent
We analysed some data sets simulated from elliptic as well as non-elliptic distributions. In each example, taking an equal number of observations from each of the two classes, we generated 500 training and test sets, each of size 200 and 500, respectively. We considered examples in dimensions 5 and 100. For classifiers based on SPD and LSPD, we used the usual sample dispersion matrix of the $j$-th ($j=1,2$) class as $\hat \sigmat_j$ when $d=5$. For $d=100$, due to statistical instability of the sample dispersion matrix, we standardized each variable in a class by its sample standard deviation.
%This is equivalent to using separate diagonal estimates of scatter matrices for the two populations.
Average test set misclassification rates of different classifiers (over 500 test sets) are reported in Table 1 along with their corresponding standard errors. To facilitate comparison, the corresponding Bayes risks are reported as well.

We compared our proposed classifiers with a pool of classifiers  %The book by Hastie et al. \cite{HTF09}, and the article by Lim et al. \cite{LLS00} contain descriptions of these classifiers and discussions about their performance. This collection of classifiers
that include parametric classifiers like LDA and QDA, and nonparametric classifiers like those based on $k$-NN (with the Euclidean metric as the distance function) and KDE (with the Gaussian kernel). For the implementation of LDA and QDA in dimension $100$, we used diagonal estimates of dispersion matrices as in the cases of SPD and LSPD. For $k$-NN and KDE, we used pooled versions of the scatter matrix estimates, which were chosen to be diagonal for $d=100$.
In Table 1, we report results for %both the single scale and
the multiscale methods of $k$-NN \cite{GCM05} and KDE \cite{GCS06} using the same weight function as described in Section 4.
% For single scale methods, the value of $k$ in $k$-NN and the bandwidth in KDE were chosen by minimizing leave-one-out cross-validation estimates of the misclassification rates \cite{HTF09}.
% For multiscale methods, we used the same weight function as described in Section 4.
%The Euclidean norm was used as the distance function in $k$-NN, and the Gaussian kernel was used in KDE.
 To facilitate comparison, we also considered SVM having the linear kernel and the radial basis function (RBF) kernel (i.e., $K_{\gamma}(\xvec,\yvec) = \exp \{-\gamma \|\xvec - \yvec\|^2 \}$ with the default value $\gamma=1/d$ as in http://www.csie.ntu.edu.tw/$\sim$cjlin/libsvm/); %artificial neural network (ANN) with a single hidden layer and the logistic transformation function, where the number of hidden nodes were chosen by minimizing a cross-validation estimate of the misclassification rate;
the classifier based on classification and regression trees (CART) and a boosted version of CART known as random forest (RF). For the implementation of SVM, CART and RF, we used the R codes available in the libraries {\tt e1071} \cite{DHLMW11}, {\tt tree} \cite{R11} and {\tt randomForest} \cite{LW02}, respectively. For classifiers based on SPD and LSPD, we wrote our own R codes using the library {\tt VGAM} \cite{YW96}, and the codes are available at {\tt https://sites.google.com/site/tijahbus/home/lspd}.

In addition, we compared the performance of our classifiers with two depth based classification methods; the classifier based on depth-depth plots (DD) \cite{LCL12}
%where the data is first transformed based on a notion of data depth and a polynomial is fitted based on minimizing the cross-validation estimate of the misclassification rate. The authors use several depth notions to construct their classifier. The misclassification rate reported here is the one corresponding to the minimum of such classifiers.
and the classifier based on maximum local depth \cite{PVB12} (LD). The DD classifier uses a polynomial of class depths (usually, half-space depth or projection depth is used, and depth is computed based on several random projections) to construct the separating surface. We used polynomials of different degrees and reported the best result in Table 1. For the LD classifier, we used the R package {\tt DepthProc} and considered the best result obtained over a range of values for the localization parameter. However, in almost all cases, the performance of the LD classifier was inferior to that of the DD classifier. So, we did not report its misclassification rates in Table 1.

%is the one corresponding to the minimum of such classifiers.%use a symmetrization idea to construct depth-based neighbors for classification purposes. Later, this idea of neighborhoods was used to develop a notion of local depth, and its inferential aspects were studied in \cite{PVB13}. We consider the maximum depth
%In this table, we put the `$\ast$' mark to indicate the best misclassification rate(s) for each data set. The other figures in bold (if any) are the misclassification rates whose differences with the best misclassification rate were found to be statistically insignificant at the 5\% level, when the usual large %sample test for proportions was used for comparison.

% All computations were carried out using a desktop computer with an {\tt Intel(R) Core(TM) i3 CPU} and {\tt 4.00 GB} of RAM. For each pair of training and test sets, the classifier based on SPD required about 3 seconds in the case $d=5$, and about 12 seconds in the case $d=100$. On the other hand, LSPD-MS took about 2 minutes and 30 seconds for $d=5$, and about 10 minutes for $d=100$.

\vspace{-.1in}
\subsection{Examples with elliptic distributions}

Recall examples {\bf E1} and {\bf E2} in Section 2 and example {\bf E3} in Section 4 involving elliptic class distributions.
%We investigate these examples to demonstrate that our proposed classifiers can be viewed as generalizations of popular parametric classifiers like LDA and QDA. Moreover, they can significantly improve upon the performance of nonparametric classifiers like those based on $k$-NN and KDE, especially in high dimensions.
In {\bf E1} with $d=5$, the DD classifier led to the lowest misclassification rate closely followed by SPD and LSPD classifiers, but in the case of $d=100$, SPD and LSPD classifiers significantly outperformed all other classifiers considered here (see Table 1). The superiority of these two classifiers was evident in {\bf E2} as well. In the case of $d=5$, the difference between their misclassification rates was statistically insignificant, though the former had an edge. Since the class distributions were elliptic, dominance of the SPD classifier over the LSPD classifier was quite expected. However, this difference was found to be statistically significant when $d=100$. In view of the normality of the class distributions, QDA was expected to have the best performance in {\bf E3}, and we observed the same. For $d=5$, the DD classifier ranked second here, while the performance of SPD and LSPD classifiers was satisfactory. However, in the case of $d=100$, SPD and LSPD classifiers again outperformed the DD classifier, and they correctly classified all the test set observations.

%\begin{center}
\begin{table*}[h]
% \begin{center}

% \hspace{-1.5in}
\renewcommand{\arraystretch}{0.7}
\caption{\small Misclassification rates (in \%) of different classifiers in simulated data sets.}
\vspace{-0.15in}
\begin{center}
{\scriptsize
%\vspace{.1in}
% \begin{tabular}{*{13}{|c}|} \hline
%\hspace{-1.0in}
\begin{tabular}{*{13}{|@{\hspace{.025mm}}c@{\hspace{.025mm}}}|} \hline

Data & ~Bayes~ & ~LDA~ & QDA & ~SVM~ & SVM & $k$-NN & KDE & ~CART~ & ~RF~ &DD & SPD & LSPD\\
set & risk & & & (LIN) & (RBF) &  & & & & & &  \\ \hline

\multicolumn{13}{|c|}{$d=5$} \\ \hline

{\bf E1} & 26.50 & 50.00 & 52.53 & 45.46 & 30.03 & 40.65 & 39.16 & 36.90 & ~31.32~ &{\bf 27.92}$\ast$ & {28.32} & {28.54} \\
& & (0.20) & (0.19) & (0.11) & (0.09) & (0.13) & (0.11) & (0.13) & ~(0.09)~ & (0.11)  & (0.10) & (0.11) \\ \hline

{\bf E2} & 0.00 & 47.43 & 42.44 & 43.92 & 38.06 & 37.64 & 34.29 & 39.10 & 34.26 & 26.68 & {\bf~ 9.26} $\ast$& {\bf 9.42} \\
& & (0.15) & (0.06) & (0.12) & (0.09) & (0.16) & (0.11) & (0.11) & (0.08) & (0.09) & (0.09) & (0.10) \\ \hline

{\bf E3} & 10.14 & 21.56 & {\bf~ 11.09} $\ast$ & 22.09 & 11.74 & 18.16 & 16.94 & 19.18 & 13.77 &{\bf 11.17} & {11.49} & 11.64 \\
& & (0.09) & (0.07) & (0.09) & (0.07) & (0.09) & (0.08) & (0.13) & (0.08) & (0.07) & (0.07) & (0.07)\\ \hline

{\bf E4} & 2.10 & 40.52 & 42.41 & 36.16 & 25.08 & {\bf 2.42} $\ast$ & {\bf 2.55} & 15.52 & 4.98 & {33.04} & 10.07 & {\bf 2.58} \\
& & (0.09) & (0.08) & (0.10) & (0.13) & (0.03) & (0.03) & (0.09) & (0.06) & (0.12) & (0.07) & (0.03)\\ \hline

{\bf E5} & 2.04 & 41.17 & 5.97 & 32.14 & 8.12 & 9.44 & 9.26 & 4.82 & {\bf 2.84} $\ast$ &5.82 & 5.65 & 5.52 \\
& & (0.15) & (0.05) & (0.34) & (0.07) & (0.08) & (0.07)& (0.08)& (0.03) & (0.05) & (0.06) & (0.06) \\ \hline

\multicolumn{13}{|c|}{$d=100$} \\ \hline

{\bf E1} & 0.48 & 50.29 & 50.67 & 46.85 & 24.97 & 44.57 & 49.99 & 35.72 & 25.14 & 24.99 & {\bf~ 1.60} $\ast$& 2.34\\
& & (0.10) & (0.13) & (0.11) & (0.06) & (0.08) & (0.10) & (0.12) & (0.12) & (0.10) & (0.11) & (0.12) \\ \hline

{\bf E2} & 0.00 & 43.77 & 46.13 & 43.99 & 40.32 & 49.96 & 49.22 & 40.30 & 32.36 & 27.56 & {\bf~ 2.90} $\ast$ & 3.18\\
& & (0.09) & (0.04) & (0.09) & (0.06) & (0.02) & (0.06) & (0.11) & (0.10) & (0.09) & (0.08) & (0.09) \\ \hline

{\bf E3} & 0.00 & 0.46 & {\bf~ 0.00} $\ast$& 3.21 & {\bf~ 0.00} $\ast$ & 49.99 & 49.98 & 17.40 & 0.02 & 1.92 & {\bf~ 0.00} $\ast$& {\bf~ 0.00} $\ast$ \\
& & (0.01) & (0.00) & (0.05) & (0.00) & (0.00) & (0.00) & (0.12) & (0.00) & (0.02) & (0.00) & (0.00)\\ \hline

{\bf E4} & 0.00 & 33.40 & 33.40 & 46.28 & 19.43 & {\bf~ 0.00} $\ast$ & {\bf~ 0.00} $\ast$ & 17.28 & {\bf~ 0.00} $\ast$ & 23.15 & {\bf~~ 0.00} $\ast$ & {\bf~~ 0.00} $\ast$ \\
& & (0.00) & (0.00) & (0.10) & (0.09) & (0.00) & (0.00) & (0.00) & (0.09) & (0.10) & (0.00) & (0.00) \\ \hline

{\bf E5} & 0.00 & 46.74 & {\bf~~ 0.00} $\ast$ & 44.45 & 7.83 & 44.01 & 49.98 & 3.32 & {\bf~~ 0.00} $\ast$ & 3.12 & {\bf~~ 0.00} $\ast$ & {\bf~~ 0.00} $\ast$ \\
& & (0.29)& (0.00)  & (0.31) & (0.15) & (0.21) & (0.04) & (0.11) & (0.00) &(0.10) & (0.00) & (0.00) \\ \hline
\end{tabular}}
 \end{center}
% \begin{center}
\vspace{-.01in}
{\footnotesize  The figure marked by `$\ast$' is the best misclassification rate observed in an example. The other figures in bold (if any) are the misclassification rates whose differences with the best misclassification rate are statistically insignificant at the 5\% level when the usual large sample test for proportion was used for comparison.
%The figure marked by `$\times$' is an example where DD classifier could not be run due to singularity.
}
% \end{center}
%\vspace{-.2in}
\end{table*}
% \end{center}

In all these examples, the Bayes classifier had non-linear class boundaries. So, LDA and SVM with linear kernel could not perform well. The performance of SVM with the RBF kernel was relatively better. In {\bf E3}, it had competitive misclassification rates for both values of $d$. $k$-NN and KDE had comparable performance in the case of $d=5$, but in the high-dimensional case ($d=100$), they misclassified almost half of the test cases. In \cite{HMN05}, the authors derived some conditions under which the $k$-NN classifier tends to classify all observations to a single class when the data dimension increases to infinity. These conditions hold in this example. It can also be shown that the classifier based on KDE with equal prior probabilities have the same problem in high dimensions. %The DD classifier did not perform well in {\bf E2} with $d=5$ because the class boundary is more complicated. For $d=100$, the DD classifier failed to yield promising results too. The performance of LD was not satisfactory, and it had misclassification rates close to 40\% in all three examples irrespective of the data dimension. So, we do not report those figures in Table 1.

\vspace{-.1in}
\subsection{Examples with non-elliptic distributions}

%Next, we considered some examples involving non-elliptic class distributions. 
Recall the trimodal example {\bf E4} discussed in Section 4. In this example, the LSPD classifier and the nonparametric classifiers based on $k$-NN and KDE significantly outperformed all other classifiers in the case of $d=5$. The differences between the misclassification rates of these three classifiers was statistically insignificant. Interestingly, along with these classifiers, the SPD classifier also led to zero misclassification rate for $d=100$. The DD classifier, LDA, QDA and SVM did not have satisfactory performance in this example.

The final example (we call it {\bf E5}) is with exponential distributions, where the component variables are independently distributed in both classes. The $i$-th variable in the first (respectively, the second) class is exponential with scale parameter $d/(d-i+1)$ (respectively, $d/2i$). Further, the second class has a location shift such that the difference between the mean vectors of the two classes is $\frac{1}{d}{\bf 1}_d=(1/d,\ldots,1/d)^{T}$. Recall that Figure 5 shows the density contours of the first class when $d=2$. In this example, the RF classifier had the best performance followed by CART when $d=5$. DD, SPD and LSPD classifiers also performed well, and their misclassification rates were significantly lower than all other classifiers. Linear classifiers like LDA and SVM with linear kernel failed to perform well. Note that as $d$ increases, the difference between the locations of these two classes shrinks to zero. This results in high misclassification rates for these linear classifiers when $d=100$. QDA performed reasonably well, and like SPD, LSPD and RF classifiers, it correctly classified all the test cases when $d=100$. The DD classifier led to an average misclassification rate of 3.12\%. Again, the classifiers based on $k$-NN and KDE had poor performance for $d=100$. This is due to the same reason as in {\bf E3} (see also \cite{HMN05}). Note that even in these examples with non-elliptic distributions, the SPD classifier performed well for high-dimensional data. This can be explained using part $(a)$ of {Theorem 4}.
%{\color{red} I am yet to decide on the version of LSPD} %The analysis of these benchmark data sets clearly shows that SPD
These examples also demonstrate that for non-elliptic or multimodal data, if not better, our LSPD classifier can perform as good as popular nonparametric classifiers. In fact, this adjustment of LSPD classifier is automatic in view of the multiscale approach developed in Section 4.

\vspace{-.1in}
\section{Analysis of benchmark data sets}

% \vspace{-.05in}
We analyzed some benchmark data sets for further evaluation of our proposed classifiers.
The biomedical data set is taken from the CMU data archive (http://lib.stat.cmu.edu/datasets/), the growth data set is obtained from \cite{RS05}, the colon data set is available in \cite{ABNGML99} (and also at the R-package `rda'), and the lightning 2 data set is taken from the UCR time series classification archive (http://www.cs.ucr.edu/$\sim$eamonn/time\_series\_data/). The remaining data sets are taken from the UCI machine learning repository (http://archive.ics.uci.edu/ml/). Descriptions of these data sets are available at these sources. In the case of biomedical data, we did not consider observations with missing values. Satellite image (satimage) data set has specific training and test samples. For this data set,
we report the test set misclassification rates of different classifiers.
If a classifier had misclassification rate $\epsilon$, its standard error was computed as $\sqrt{\epsilon(1-\epsilon)/\mbox{(the size of the test set)}}$. In all other data sets, we formed the training and the test sets by randomly partitioning the data, and this random partitioning was repeated 500 times to generate new training and test sets. The average test set misclassification rates of different classifiers are reported in Table 2 along with their corresponding standard errors. The sizes of the training and the test sets in each partition are also reported in this table.
Since the codes for the DD classifier are available only for two class problems, we could use it only in cases of biomedical and Parkinson's data, where it yielded misclassification rates of 12.54\% and 14.48\%, respectively, with corresponding standard error of 0.18\% and 0.15\%. In the case of growth data, where training sample size from each class was smaller than the dimension, the values of randomized versions of half-space depth and projection depth were zero for almost all observations. Due to this problem, the DD classifier could not be used. We used the maximum LD classifier on these real data sets, but in most of the cases, its performance was not satisfactory. So, we do not report them in Table 2.

\begin{table*}[htp]
\renewcommand{\arraystretch}{.7}
% \begin{center}
\caption{\small Misclassification rates (in \%) of different classifiers in real data sets.}

{\scriptsize
\begin{center}
\vspace{-.1in}
\begin{tabular}{*{7}{|@{\hspace{.01mm}}c@{\hspace{.01mm}}}|*{3}{|@{\hspace{.01mm}}c@{\hspace{.01mm}}}|} \hline
% \begin{tabular}{|c|c|c|c|c|c|c||c|c|c|} \hline

{Data set} & {Biomed} & Parkinson's & Wine & Waveform & Vehicle & Satimage & Growth &  Lightning 2 & Colon \\ \hline \hline

Dimension ($d$) & 4 & 22 & 13 & 21 & 18 & 36 & 31 &  637 & 2000 \\
Classes ($J$) & 2 & 2 & 3 & 3 & 4 & 6 & 2 & 2 & 2 \\ \hline
Training size & 100 & 97 & 100 & 300 & 423 & 4435 & 46  & 60 & 31 \\
Test size & 94 & 98 & 78 & 501 & 423 & 2000 & 47 &  61 & 31 \\ \hline \hline

LDA & 15.66 & 30.93 & 2.00 & 19.90 & 22.49 & 16.06 & 29.15  & 32.51 & {\bf~~ 14.03} $\ast$ \\
& (0.14) & (0.12) & (0.06) & (0.15) & (0.07) & (0.82) & (0.34)  & (0.35) & (0.20) \\

QDA & {\bf 12.57} & xxxx & 2.46 & 21.12 & {\bf 16.38} & 14.14 & xxxx & xxxx  & xxxx \\
& (0.13) & xxxx & (0.09) & (0.15) & (0.07) & (0.78) & xxxx & xxxx & xxxx \\ \hline

SVM (LIN) & 22.03 & 15.31 & 3.64 & 18.88 & 21.20 & 15.18 & 5.16  & 35.64 & 16.38 \\
& (0.13) & (0.12) & (0.09) & (0.07) & (0.07) & (0.80) & (0.12) &  (0.35) & (0.23) \\

SVM (RBF) & {\bf 12.76} & 13.69 & {1.86} & 16.08 & 25.57 & 30.99 & {\bf~~ 4.66} $\ast$  & 28.73 & 35.48 \\
& (0.13) & (0.10) & (0.06) & (0.07) & (0.08) & (1.03) & (0.11) & (0.32) & (0.00) \\ \hline

% $k$-NN & 18.00 & 14.54 & 2.04 & 17.04 & 21.84 & {\bf~~ 9.23} $\ast$& {\bf 4.90} & {30.04} & 23.03 \\
% & (0.15) & (0.16) & (0.07) & (0.08) & (0.08) & (0.65) & (0.12) & (0.23) & (0.28) \\

$k$-NN  & 17.74 & 14.42 & 1.98 & 16.37 & 21.80 & {\bf~~ 9.23} $\ast$& {\bf 4.48} & 30.09 & 22.47 \\
& (0.15) & (0.16) & (0.06) & (0.08) & (0.08) & (0.65) & (0.10) & (0.20) & (0.27) \\

% KDE & 16.93 & {\bf 11.09} & {1.69} & 24.40 & 21.45 & 19.81 & 5.30 & {28.17} & 23.42 \\
% & (0.15) & (0.12) & (0.06) & (0.03) & (0.07) & (0.89) & (0.13) &  (0.31) & (0.28) \\

KDE  & 16.67 & {\bf~~ 11.01 $\ast$} & {\bf~~ 1.36} $\ast$ & 23.83 & 21.21 & 19.81 & {\bf 4.79} & 28.11 & 23.20 \\
& (0.14) & (0.12) & (0.05) & (0.03) & (0.07) & (0.89) & (0.13) & (0.30) & (0.28) \\ \hline

% ANN & 25.85 & 24.39 & 6.12 & 55.11 & 23.97 & 14.55 & 13.30 &  30.50 & 22.58 \\
% & (0.19) & (0.07) & (0.27) & (0.12) & (0.11) & (0.79) & (0.27) & (0.29) & (0.30) \\

CART & 17.69 & 16.63 & 10.99 & 56.61 & 31.41 & 53.43 & 17.40 & 33.96 & 28.78 \\
& (0.18) & (0.20) & (0.22) & (0.12) & (0.10) & (0.56) & (0.25) & (0.34) & (0.35) \\

RF & 13.23 & 11.58 & 2.12 & 57.02 & 25.52 & 30.91 & 9.67 & {\bf 22.08}$\ast$ & 19.10 \\
& (0.14) & (0.15) & (0.06) & (0.12) & (0.07) & (0.48) & (0.25) & (0.34) & (0.30) \\ \hline

SPD & {\bf 12.53} & 15.44 & 2.34 & {\bf~~ 15.12} $\ast$ & {\bf~~ 16.35} $\ast$ & 12.59 & 14.64 & {27.70} & 19.98 \\
& (0.21) & (0.15) & (0.08) & (0.06) & (0.08) & (0.74) & (0.28) &  (0.30) & (0.31) \\

LSPD & {\bf~~ 12.49} $\ast$& {\bf 11.35} & {1.85} & {\bf 15.36} & 17.15 & 12.59 & 5.10 & {27.46}  & 20.51 \\
& (0.15) & (0.11) & (0.07) & (0.06) & (0.08) & (0.74) & (0.12) &(0.30) & (0.33) \\ \hline

\end{tabular}
%\end{center}
\end{center}}
%\vspace{.1in}
% \begin{center}
{\footnotesize
The figure marked by `$\ast$' is the best misclassification rate observed for a data set. The other figures in bold (if any) are the misclassification rates whose differences with the best misclassification rate are statistically insignificant at the 5\% level. Because of the singularity of the estimated class dispersion matrices, QDA could note be used in some cases and those are marked by `xxxx'.}
% \end{center}
\vspace{-.1in}
\end{table*}

In biomedical and vehicle data sets, scatter matrices of the competing classes were very different. So, QDA had significant improvement over LDA. In fact, its misclassification rates of QDA were close to the best ones. In both of these data sets, the class distributions were nearly elliptic
(this can be verified using the diagnostic plots suggested in \cite{LFZ97}). The SPD classifiers utilized the ellipticity of the class distributions to outperform the nonparametric classifiers. The LSPD classifier could compete with the SPD classifier in the biomedical data. But in the vehicle data, where the evidence of ellipticity was much stronger, it had a slightly higher misclassification rate.
% (not shown in this article)

In the Parkinson's data set, we could not use QDA because of the singularity of the estimated class dispersion matrices. So, we used the estimated pooled dispersion matrix for standardization in our classifiers. In this data set, all nonparametric classifiers had significantly lower misclassification rates than LDA. Among them, the classifier based on KDE had the lowest misclassification rate. The performance of LSPD classifier was also competitive. Since the underlying distributions were non-elliptic, the LSPD classifier significantly outperformed the SPD classifier. We observed almost the same phenomenon in the wine data set as well, where the classifier based on KDE yielded the best misclassification rate followed by the LSPD classifier. In these two data sets, although the data dimension was quite high, all competing classes had low intrinsic dimensions (can be estimated using \cite{LB05}). So, the nonparametric methods like KDE were not much affected by the curse of dimensionality. Recall that for small values of $h$, LSPD$_h$ performs like KDE. Therefore, the difference between the misclassification rates of KDE and LSPD classifiers was statistically insignificant.

In the waveform data set, the SPD classifier had the best misclassification rate. %closely followed by the LSPD classifier. %The pairwise scatter plot of ${\hat{\zvec}(\xvec_{ij})}$s showed that the estimated SPD contained sufficient information about class separability. Further, the diagnostic plots \cite{LFZ97} showed that 
In this data set, the class distributions were nearly elliptic. So, the SPD classifier was expected to perform well. As the LSPD classifier is quite flexible, it yielded competitive misclassification rates. Here, the class distributions were not normal (can be checked using the method in \cite{R83}), and they did not have low intrinsic dimensions. As a result, other parametric as well as nonparametric classifiers had relatively higher misclassification rates.

Recall that in the satimage data set, results are based on a single training and a single test set. So, the standard errors of the misclassification rates were high for all classifiers, and this makes it difficult to compare the performance of different classifiers. In this data set, $k$-NN classifiers led to the lowest misclassification rate, but SPD and LSPD classifiers performed better than all other classifiers. Nonlinear SVM, CART and RF had quite high misclassification rates.
% of the classifiers based on KDE was also quite high. %This was due to the poor choice of the associated tuning parameter.

We further analyzed some data sets, where the sample size was quite small compared to data dimension. In these data sets, we worked with unstandardized observations. Instead of using the estimated pooled dispersion matrix, we used the identity matrix for implementation of LDA. The growth data set contains growth curves of males and females, which are smooth and monotonically increasing functions. Because of high dependence among the measurement variables, the class distributions had low intrinsic dimensions, and they were non-elliptic. As a result, the nonparametric classifiers performed well. SVM with the RBF kernel had the best misclassification rate, but those of $k$-NN, KDE and LSPD classifiers were also comparable. Good performance of the linear SVM classifier indicates that there was a good linear separability between the two classes, but LDA failed to figure it out.

The lightning 2 data set consists of observations that are realizations of time series.  %For this data set, the authors in \cite{EHDPMPT02} reported good performance of the classifier based on SVM with the RBF kernel. But,
In this data set, RF had the best performance followed by the LSPD classifier. Here also, we observed non-elliptic class distributions with low intrinsic dimensions \cite{LB05}. This justifies the good performance of the classifiers based on $k$-NN and KDE.
%In this example, the LSPD-MS classifiers had lowesr misclassification rates than classifiers based on KDE and $k$-NN.
The SPD classifier also had competitive misclassification rates because of the flexibility of GAM. In fact, it yielded the third best performance in this data set.

Finally, we analyzed the colon data set, which contains micro-array expressions of 2000 genes for some `normal' and `colon cancer' tissues. In this data set, there was good linear separability among the observations from the two classes. So, LDA and linear SVM had lower misclassification rates than all other classifiers. Among the nonparametric classifiers, RF had the best performance closely followed by the SPD classifier. These two classifiers were less affected by the curse of dimensionality. Recall that LSPD$_h$ with large bandwidth $h$ approximates SPD. Because of this automatic adjustment, the LSPD classifier could nearly match the performance of the SPD classifier. 
%As a matter of fact, the misclassification rates of these classifiers were significantly lower than all other nonlinear classifiers considered here.

\begin{figure}[t]
\begin{center}
% \hspace{-1in}
{\includegraphics[height=7.0cm, width=14.0cm]{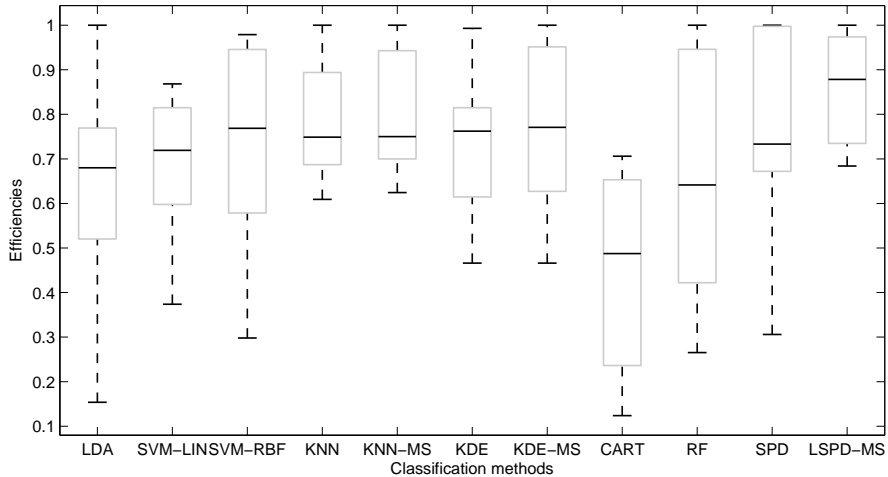}}
\end{center}
\vspace{-.35 in}
\caption{\tt Overall efficiencies of different classifiers.}
\vspace{-.1 in}
\end{figure}

To compare the overall performance of different classifiers, following the idea of \cite{CGO09, F96}, we computed their efficiency scores on different data sets. For a data set, if $T$ classifiers have misclassification rates $\epsilon_1, \ldots, \epsilon_T$, the efficiency of the $t$-th classifier ($e_t$) is defined as $e_t = \epsilon_0/\epsilon_t$, where $\epsilon_0 = \min_{1\le t\le T} \epsilon_t$. Clearly, in any data, the best classifier has $e_t=1$, while a lower value of $e_t$ indicates the lack of efficiency of the $t$-th classifier. In each of these benchmark data sets, we computed this ratio for all classifiers, and they are graphically represented by box plots in Figure 7. This figure clearly shows the superiority of the LSPD classifier (with the highest median value of 0.88) over its competitors. We did not consider QDA for comparison because it could not be used for some of the data sets.

\vspace{-.1in}
\section{Concluding remarks}

In this article, we develop and study classifiers constructed by fitting a nonparametric additive logistic regression model to features extracted from the data using SPD as well as its localized version, LSPD. The SPD classifier can be viewed as a generalization of parametric classifiers like LDA and QDA. When the underlying class distributions are elliptic, it has Bayes optimality. For large values of $h$, while LSPD$_h$ behaves like SPD, for small values of $h$, it captures the underlying density. So, the multiscale classifier based on LSPD is flexible, and it overcomes several drawbacks associated with SPD and other existing depth based classifiers. When the underlying class distributions are elliptic but not normal, both SPD and LSPD classifiers outperform popular parametric classifiers like LDA and QDA as well as nonparametric classifiers. In the case of non-elliptic or multi-modal distributions, while SPD may fail to extract meaningful discriminatory features, the LSPD classifier can compete with
other nonparametric methods. Moreover, for high-dimensional data, while traditional nonparametric methods suffer from the curse of dimensionality, both SPD and LSPD classifiers can lead to low misclassification probabilities. %. For distributions in high dimensions (elliptic as well as non-elliptic), they may yield significantly better performance compared to
%and they can outperform popular nonparametric classifiers like those based on KDE and $k$-NN. %In particular, this holds when the scale difference between the two classes dominates the difference in their locations. We observed this in the simulated examples analyzed in Section 6, and it can be explained using the theoretical results stated in {Theorem 3} and \cite{HMN05}. 
Analyzing several simulated and benchmark data sets, we have amply demonstrated this. %in this article.
%Note that %The LSPD classifier had the best overall performance on benchmark data sets (see Figure 7). In 
%in biomedical, vehicle and waveform data sets, where the class distributions were nearly elliptic, SPD and LSPD classifiers performed better than all nonparametric classifiers considered here. For Parkinson's and wine data, where the class distributions were non-elliptic, SPD classifier failed, but the LSPD classifier had misclassification rates close to that of the best nonparametric classifier. 
%Unlike simulated examples, 
In high-dimensional benchmark data sets, the class distributions had low intrinsic dimensions due to high correlation among the the measurement variables \cite{LB05}. Moreover, the competing classes differed mainly in their locations. As a consequence, though the proposed LSPD classifier had the best overall performance in benchmark data sets, its superiority over other nonparametric methods was not as prominent as it was in the simulated examples.

%\vspace{-.1in}
%\section*{Acknowledgement}

%The authors are thankful to Prof. Probal Chaudhuri for his valuable contributions to this manuscript.
%They would also like to thank Dr. Thomas W. Yee for his help with VGAM, and Dr. Jun Li for sharing the codes of DD classifier.
% The authors would like to thank the associate editor and the referees for their careful reading of the earlier version of the manuscript and providing several helpful comments.

% \appendix
\vspace{-.15in}
\section*{Appendix : Proofs and Mathematical Details}

\noindent
{\bf Lemma 1 :} If $F$ has a spherically symmetric density $f(\xvec)=g(\|\xvec\|)$ on $\mathbb{R}^d$ with $d>1$, then $\|E_F[u(\xvec-\Xvec)]\|$ is a non-negative monotonically increasing function of\;$\|\xvec\|$.
\vspace{.05in}

\noindent
{\bf Proof of {Lemma 1} :} In view of spherical symmetry of $f(\xvec)$, $S(\xvec)=\|E_F[u(\xvec-\Xvec)]\|$ is invariant under orthogonal transformations of $\xvec$. Consequently, $S(\xvec)=\eta(\|\xvec\|)$ for some non-negative function $\eta$. Consider now $\xvec_1$ and $\xvec_2$ such that $\|\xvec_1\| < \|\xvec_2\|$.
Using spherical symmetry of $f(\xvec)$, without loss of generality, we can assume $\xvec_i=(t_i,0,\ldots,0)^{T}$ for $i=1,2$ such that $|t_1| < |t_2|$. For any $\xvec=(t,0,\ldots,0)^{T}$, we have
$$S(\xvec) = \biggl| E_F \biggl [ \frac{(t-X_1)}{\sqrt{(t-X_1)^2 + X_2^2 + \ldots + X_d^2}} \biggr ] \biggr|,$$
due to spherical symmetry of $f(\xvec)$.
Note also that for any $\xvec \in \mathbb{R}^d$ with $d>1$, $E_F[\|\xvec - \Xvec\|]$ is a strictly convex function of $\xvec$ in this case. Consequently, it is a strictly convex function of $t$. %when $\xvec=(t,0,\ldots,0)^{'}$.
Observe now that $S(\xvec)$ with this choice of $\xvec$ is the absolute value of the derivative of $E_F[\|\xvec - \Xvec\|]$ w.r.t. $t$. This derivative is a symmetric function of $t$ that vanishes at $t=0$. Hence, $S(\xvec)$ is an increasing function of $|t|$, and this proves that $\eta(\|\xvec_1\|) < \eta(\|\xvec_2\|)$. $\hfill \Box$

\vspace{.05in}
\noindent
{\bf Proof of {Theorem 1} :}
% Recall the form of $f_i$ described before the statement of {Theorem 1}.
%For $1 \leq i \leq J$, consider the Mahalanobis distance,
%$\delta(\xvec, F_i) = \{(\xvec - \muvec_{i})^{\prime}$ $\sigmat_{i}^{-1} (\xvec - \muvec_{i})\}^{1/2} = \|\sigmat_i^{-1/2}(\xvec - \muvec_i)\|$.
If the population distribution $f_j(\xvec)$ ($1 \leq j \leq J$) is elliptically symmetric, we have $f_j(\xvec) = |\sigmat_j|^{-1/2} g_j({\delta}(\xvec, F_j))$, where $\delta(\xvec, F_j) = \{(\xvec - \muvec_{j})^{T}$ $\sigmat_{j}^{-1} (\xvec - \muvec_{j})\}^{1/2}=\|\sigmat_j^{-1/2}(\xvec - \muvec_j)\|$. Since SPD$(\xvec,F_j)=1-\|E\{u(\sigmat_{j}^{-1} (\xvec - \muvec_{j}))\}\|$  is affine invariant, it is a function of $\delta(\xvec,F_j)$, the Mahalanobis distance. Again, since $\sigmat_j^{-1/2}(\Xvec - \muvec_j)$ has a spherically symmetric distribution with its center at the origin, from {Lemma 1} it follows that SPD$(\xvec,F_j)$ is a monotonically decreasing function of $\delta(\xvec,F_j)$. So, $\delta(\xvec,F_j)$ is also a function of SPD$(\xvec,F_j)$. Therefore, $f_j(\xvec)$, which is a function of $\delta(\xvec,F_j)$, can also be expressed  as
$$f_j(\xvec) = \psi_j(\mbox{SPD}(\xvec,F_j))~\mbox{for all}~ 1 \leq j \leq J,$$
where $\psi_j$ is an appropriate real-valued function that depends on $g_j$. Now,
one can check that
$$\log \{p(j|\xvec) / p(J|\xvec)\}=\log(\pi_j/\pi_J) + \log \psi_j(\mbox{SPD}(\xvec,F_j)) - \log \psi_J(\mbox{SPD}(\xvec,F_J)).$$
for $1 \leq j \leq (J-1)$. Now, for $1 \leq j \neq i \leq (J-1)$, define  $\varphi_{jj}(z) = \log \pi_j + \log \psi_j(z)$ and $\varphi_{ij}(z)=0$.
So, if we define $\varphi_{1J}(z)=\ldots=\varphi_{(J-1)J}(z) = - \log \pi_J - \log \psi_J(z)$, the proof of the theorem is complete. $\hfill \Box$

\vspace{.05in}
\noindent
{\bf Remark 1} : If $f_j(\xvec)$ is unimodal, $\psi_j(z)$ is monotonically increasing for $1 \leq j \leq J$. Moreover, if the distributions differ only in their locations, the $\psi_j(z)$s are same for all class. In that case, $f_j(\xvec)>f_i(\xvec) \Leftrightarrow \delta(\xvec,F_j)>\delta(\xvec,F_i)$ for $1 \leq i \neq j \leq J$, and hence the classifier turns out to be the maximum depth classifier.

\vspace{.05in}
\noindent
{\bf Proof of {Theorem 2 (a)} :} Let $h<1$. For any fixed $\xvec \in \mathbb{R}^d$ and the distribution function $F_j$, we have
$\mbox{LSPD}_h(\xvec,F_j) = E_{F_j}[K_h(\tvec)] - \| E_{F_j} [K_{h}({\tvec}) u(\tvec)] \|,$
where $\tvec = {\sigmat_{j}}^{-1/2}(\xvec - \Xvec)$ for $1 \leq j \leq J$.
For the first term in the expression of LSPD$_h(\xvec,F_j)$ above, we have
\begin{eqnarray*}
E_{F_j}[K_h(\tvec)] = \int_{\mathbb{R}^d} {h}^{-d} K_{h}(\sigmat_{j}^{-1/2}(\xvec-\vvec)) f_j(\vvec) d\vvec = |\sigmat_{j}|^{1/2} \int_{\mathbb{R}^d} K(\yvec) f_j(\xvec-h\sigmat_{j}^{1/2}\yvec) d\yvec, \nonumber
\end{eqnarray*}
\noindent
{where} $\yvec = h^{-1}\sigmat_{j}^{-1/2}(\xvec -\vvec)$. So, using Taylor's expansion of $f_j(\xvec)$, we get
\begin{eqnarray*}
E_{F_j}[K_h(\tvec)] = |\sigmat_{j}|^{1/2} f_j(\xvec) - h |\sigmat_{j}|^{1/2} \int_{\mathbb{R}^d} K(\yvec) ~(\sigmat_{j}^{1/2}\yvec)^{\prime} \nabla f_j(\xivec) d\yvec, \nonumber
% && ~ \mbox{value theorem, and that}~K(\yvec)~\mbox{is a density function}). \nonumber
\end{eqnarray*}
%\setlength{\arraycolsep}{5pt}
%The last equality above follows by Taylor's expansion of $f_j$,
where $\xivec$ lies on the line joining $\xvec$ and $(\xvec-h\sigmat_{j}^{1/2}\vvec)$. So, using the Cauchy-Scawartz inequality, one gets
%\begin{equation}
$\Bigl| E_{F_j}[K_h(\tvec)] - |\sigmat_{j}|^{1/2} f_j(\xvec)\Bigr| \leq h |\sigmat_{j}|^{1/2} \lambda_{j}^{1/2} M^{\circ}_{j} M_K$,
%\end{equation}
where $M^{\circ}_{j}=\sup_{\xvec \in \mathbb{R}^d} \|\nabla f_j(\xvec)\|$, $M_K = \int \|\yvec\| K(\yvec) d\yvec$, and $\lambda_{j}$ is the largest eigenvalue of $\sigmat_{j}$ for $1 \leq j \leq J$.
This implies $\Bigl| E_{F_j}[K_h(\tvec)] - |\sigmat_{j}|^{1/2} f_j(\xvec)\Bigr| \rightarrow 0$ as $h \rightarrow 0$ for $1 \leq j \leq J$.

%\noindent
For the second term in the expression of LSPD$_h(\xvec,F_j)$, a similar argument yields
\begin{eqnarray*}
&& E_{F_j} [K_{h}({\tvec}) u(\tvec)] = |\sigmat_{j}|^{1/2} \int_{\mathbb{R}^d} K(\yvec) u({\yvec}) f_j(\xvec-h\sigmat_{j}^{1/2}\yvec) d\yvec~~~~~~~~~~~~~~~~~~~~~~~ \nonumber \\
&& ~~~~~~~~~~~~~~~~= - h |\sigmat_{j}|^{1/2} \int_{\mathbb{R}^d} K(\yvec) u({\yvec}) ~(\sigmat_{j}^{1/2}\yvec)^{\prime} \nabla f_j(\xivec) d\yvec
~~(\mbox{since} \int K(\yvec)u(\yvec)d\yvec={\bf 0}). \nonumber
% && ~ \mbox{value theorem, and the fact} ~K(\yvec)u({\yvec}) ~\mbox{is an odd function}). \nonumber
\end{eqnarray*}
So,
%\begin{equation}
$\|E_{F_j}[K_{h}({\tvec}) u(\tvec)]\| \leq h |\sigmat_{j}|^{1/2} \lambda_{j}^{1/2} M^{\circ}_{j} M_K \rightarrow 0~\mbox{as} ~h~ \rightarrow 0$, and
%\end{equation}
%(6) and (7) together imply that
hence, $\mbox{LSPD}_h(\xvec,F_j) \rightarrow |\sigmat_{j}|^{1/2} f_j(\xvec) ~\mbox{as}~ h \rightarrow 0.$
Consequently,
$\zvec_h(\xvec)%($LSPD$_h(\xvec, F_1)$, $\ldots$, LSPD$_h(\xvec, F_J))
\rightarrow (|\sigmat_{1}|^{1/2} f_1(\xvec), \ldots, |\sigmat_{J}|^{1/2}f_J(\xvec))^T$ as $h \rightarrow 0$.
\hfill $\Box$

\vspace{.05in}
\noindent
{\bf Proof of {Theorem 2 (b)} :} Here we consider the case $h>1$. Consider any fixed $\xvec \in \mathbb{R}^d$ and any fixed $j$ ($1 \leq j \leq J$). For any fixed $\tvec$, since
$K(\tvec/h) \rightarrow K({\bf 0})$ as $h \rightarrow \infty$, using Dominated Convergence Theorem (note that $K$ is bounded), one can show that
$${\mbox{LSPD}}_h(\xvec,F_j) \rightarrow K({\bf 0}) {\mbox{SPD}}(\xvec,F_j)~\mbox{as}~h \rightarrow \infty.$$
%$h>1$) in view of the boundedness of the kernel function.
So,
$\zvec_h(\xvec)%=($LSPD$_h(\xvec,$ $F_1)$, $\ldots$, LSPD$_h(\xvec, F_J))
\rightarrow %\linebreak
(K({\bf 0})$SPD$(\xvec, F_1), \ldots, K({\bf 0})$SPD$(\xvec, F_J))^T$ as $h \rightarrow \infty$. \hfill $\Box$

\vspace{.05in}
\noindent
{\bf Proof of Theorem 3 :} Define the sets $B_n=\{ \xvec=(x_1,\ldots,x_d) : \|\xvec\| \leq \sqrt{d}n \}$, and $A_n=\{\xvec:~n^2x_i ~\mbox{is an integer and}~ |x_i|\leq n ~\mbox{for all}~ 1 \leq i \leq d \}$. Clearly $A_n \subset B_n \subset \mathbb{R}^d$, the set $B_n$ is a closed ball and the set $A_n$ has cardinality $(2n^3+1)^d$.
We will prove the almost sure (a.s.) uniform convergence on the three sets: (i) on $A_n$ (ii) on $B_n \setminus A_n$, and (iii) on $B_n^c$.
%We first consider the set $A_n$.

\noindent
Consider any fixed $h \in (0,1]$. Recall that for this choice of $h$, LSPD$_h(\xvec,F)$ (see equation (3)) and LSPD$_h(\xvec,F_n)$ are defined as follows:
$$\mbox{LSPD}_h(\xvec,F_{n}) = \frac{1}{nh^d} \sum_{i=1}^n K(h^{-1}({\xvec - \Xvec_i})) - \left\|\frac{1}{nh^d} \sum_{i=1}^n K(h^{-1}({\xvec - \Xvec_i})) ~u(\xvec - \Xvec_i)\right\|,$$
$$ \mbox{and}~ \mbox{LSPD}_h(\xvec,F) = h^{-d}E[K(h^{-1}({\xvec - \Xvec}))] - h^{-d}\|E[K(h^{-1}({\xvec - \Xvec})) ~u(\xvec - \Xvec)]\|.$$

(i) Define $ \Zvec_i = K(h^{-1}({\xvec - \Xvec_i})) u(\xvec - \Xvec_i) - E[K(h^{-1}({\xvec - \Xvec})) u(\xvec - \Xvec)]$ for $1 \leq i \leq n$. Note that $\Zvec_i$s are independent and identically distributed (i.i.d.) with $E(\Zvec_i)={\bf 0}$ and $\|\Zvec_i\| \le 2K({\bf 0})$. Using the exponential inequality for sums of i.i.d. random vectors (see p. 491 of \cite{Y76}), for any fixed $\epsilon>0$, we get
$P\Bigl(\| \frac{1}{n} \sum_{i=1}^{n} \Zvec_i \|\ge \epsilon \Bigr) \leq 2 e^{-C_0n\epsilon^2}$,
where $C_0$ is a positive constant that depends on $K({\bf 0})$ and $\epsilon$.
This now implies that
%\vspace{0.05in}
{\[
\begin{array}{l}
\vspace{0.1in}
P\biggl(\Bigl\|\frac{1}{nh^d} \sum_{i=1}^{n}K(h^{-1}({\xvec - \Xvec_i})) u(\xvec - \Xvec_i)\Bigr\| - \Bigl\|h^{-d}E[K(h^{-1}({\xvec - \Xvec})) u(\xvec - \Xvec)] \Bigr\| \ge \epsilon\biggr) \\
\vspace{-0.05in}
\leq P\biggl(\Bigl\|\frac{1}{nh^d} \sum_{i=1}^{n}K(h^{-1}({\xvec - \Xvec_i})) u(\xvec - \Xvec_i) - h^{-d}E[K(h^{-1}({\xvec - \Xvec})) u(\xvec - \Xvec)] \Bigr\| \ge \epsilon\biggr)\end{array} \]}
\vspace{-0.1in}
\begin{equation}
\hspace{-3.25in}= P\Bigl(\| \frac{1}{n} \sum_{i=1}^{n} \Zvec_i \|\ge h^{d}\epsilon \Bigr) \leq 2 e^{-C_0nh^{2d}\epsilon^2}.
\vspace{-0.05in}
\end{equation}
For a fixed value of $h$, since $\sum_{i=1}^n K(h^{-1}({\xvec - \Xvec_i}))$ is a sum of i.i.d. bounded random variables, using Bernstein's inequality, we also have
\vspace{-0.1in}
$$P\biggl(\Bigl|\frac{1}{n} \sum_{i=1}^n K(h^{-1}({\xvec - \Xvec_i}))-E[K(h^{-1}({\xvec - \Xvec}))]\Bigr|\ge \epsilon \biggr) \le 2 e^{-C_1n\epsilon^2}$$ for some suitable positive constant $C_1$. This implies
\vspace{-0.1in}
\begin{equation}
P\biggl(\Bigl|\frac{1}{nh^d} \sum_{i=1}^n K(h^{-1}({\xvec - \Xvec_i}))- h^dE[K(h^{-1}({\xvec - \Xvec}))]\Bigr|\ge \epsilon\biggr) \le 2 e^{-C_1nh^{2d}\epsilon^2}.
\vspace{-0.1in}
\end{equation}
Combining equations (6) and (7), we get
%\vspace{-0.1in}
$ P(|\mbox{LSPD}(\xvec,F_n)-\mbox{LSPD}(\xvec,F)|\ge \epsilon) \le  C_3 e^{-C_4nh^{2d}\epsilon^2}$
%\vspace{-0.05in}
for some suitable constants $C_3$ and $C_4$. Since the cardinality of $A_n$ is $(n^3+1)^d$, we have
\vspace{-0.1in}
\begin{equation}
P(\sup_{\xvec \in A_n}|\mbox{LSPD}(\xvec,F_n)-\mbox{LSPD}(\xvec,F)|\ge \epsilon)\le C_3 (n^3+1)^de^{-C_4nh^{2d}\epsilon^2}.
\vspace{-0.1in}
\end{equation} 
%Hence, for any fixed $h\le 1$,
Now, $\sum_{n\ge 1} (n^3+1)^de^{-C_4nh^{2d}\epsilon^2} < \infty$. So, a simple application of Borel-Cantelli lemma implies that
$\sup_{\xvec \in A_n}|\mbox{LSPD}_h(\xvec, F_n)-\mbox{LSPD}_h(\xvec, F)| \stackrel{a.s.}{\rightarrow} 0~\mbox{as}~n \to \infty.$
%$$P(\sup_{\xvec \in A_n} \| \overline{\Yvec}_{h}(\xvec) - E\{\overline{\Yvec}_{h}(\xvec)\} \| > \epsilon) \leq 2n^{3d} e^{-2nh^{2d}\epsilon^2/M_K^2}.$$
%This now implies that there exists a constant $C_1>0$ such that
%\begin{equation}
%\sup_{\xvec \in A_n} \| \overline{\Yvec}_{h}(\xvec) - E\{\overline{\Yvec}_{h}(\xvec)\} \| \leq C_1 \sqrt{{\log n}/{nh^{2d}}}.
%\end{equation}

%\vspace{.05 in}
(ii) Consider the set $B_n \setminus A_n$.
%$$\sup_{\xvec \in B_n \setminus A_n} \| \overline{\Yvec}_{h}(\xvec) - E\{\overline{\Yvec}_{h}(\xvec)\} \| \leq \sup_{\xvec \in B_n \setminus A_n} \| \overline{\Yvec}_{h}(\xvec) \| + h M_f^{\prime} I_K.$$
Note that given any $\xvec$ in $B_n \setminus A_n$, there exists $\yvec \in A_n$ such that $\|\xvec - \yvec\| \leq {\sqrt{2}}/{n^2}$. First we will show that
$|\mbox{LSPD}(\yvec, F_n) - \mbox{LSPD}(\xvec, F_n)| \stackrel{a.s.}{\rightarrow} 0$ as $n \rightarrow \infty$.
%For any fixed $h\le1$,
Using the mid-value theorem, one gets
\vspace{-0.1in}
{\small \[\left|\frac{1}{nh^d} \sum_{i=1}^{n} K(h^{-1}(\xvec -\Xvec_i)) - \frac{1}{nh^d} \sum_{i=1}^{n} K(h^{-1}(\yvec -\Xvec_i))\right| \leq \frac{1}{nh^{d+1}} \sum_{i=1}^{n} \bigl |(\xvec-\yvec)^{T} \nabla K[(\xivec-\Xvec_i)/h] \bigr |,
\vspace{-0.1in}
\]}
where $\xivec$ lies on the line joining $\xvec$ and $\yvec$. Note that the right hand side is less than $\frac{M_{K}^{'}}{h^{d+1}}\frac{\sqrt{2}}{n^2}$, where
$M_{K}^{'}= \sup_{\tvec} \|\nabla K(\tvec)\|$.
This upper bound is free of $\xvec$, and
%. Since $M_{K}^{'}<\infty$, $\frac{M_{K}^{'}}{h^{d+1}}\frac{\sqrt{2}}{n^2}
goes to $0$ as $n \rightarrow \infty$.
Now,
\vspace{-0.1in}
$$ \left\| \frac{1}{nh^d} \sum_{i=1}^{n} K(h^{-1}(\xvec-\Xvec_i)) u(\xvec- \Xvec_i)\right\| - \left\|\frac{1}{nh^d} \sum_{i=1}^{n} K(h^{-1}(\yvec-\Xvec_i)) u(\yvec- \Xvec_i)\right\|
\vspace{-0.1in}
$$
$$\le \left\| \frac{1}{nh^d} \sum_{i=1}^{n} [ K(h^{-1}(\xvec-\Xvec_i)) u(\xvec- \Xvec_i) - K(h^{-1}(\yvec-\Xvec_i)) u(\yvec- \Xvec_i)] \right\|
\vspace{-0.1in}
$$
$$\le \left| \frac{1}{nh^d} \sum_{i=1}^{n} [K(h^{-1}(\xvec-\Xvec_i)) - K(h^{-1}(\yvec-\Xvec_i))] \right|
+ K({\bf 0}) \left\| \frac{1}{nh^d} \sum_{i=1}^n  \{u(\xvec - \Xvec_i) - u(\yvec - \Xvec_i)\} \right \|.$$
% $$\small \leq {M_K}{h^{-d}} \sup_{\xvec \in B_n \setminus A_n} \| {n}^{-1} \sum_{i=1}^n [u(\xvec - \Xvec_i) - u(\yvec - \Xvec_i)] \| + {M_K^{\prime}\|\xvec - \yvec\|}{h^{-(d+1)}}.$$
%$$\leq {M_K}{h^{-d}} \sup_{\xvec \in B_n \setminus A_n} \bigg \|{n}^{-1} \sum_{i=1}^n \{u(\xvec - \Xvec_i) - u(\yvec %- \Xvec_i)\} \bigg \| + {2 M_K^{\prime}n^{-2}}{h^{-(d+1)}}.$$

We have already proved that the first part converges to $0$ in a.s. sense.
For the second part, consider a ball of radius $1/n$ around $\xvec$ (say, $B(\xvec, 1/n)$). Now,
\vspace{-0.05in}
\[ K({\bf 0}) \left\| \frac{1}{nh^d}\sum_{i=1}^n \{u(\xvec - \Xvec_i) - u(\yvec - \Xvec_i)\} \right \| \leq \bigg | \frac{2K({\bf 0})}{nh^d} \sum_{i=1}^n I\{\Xvec_i \in B(\xvec, 1/n)\} \bigg | + \frac{2nK({\bf 0})}{h^d}\|\xvec - \yvec\|
\vspace{-0.1in}
\]
$$ \le \frac{2K({\bf 0})}{h^d} \bigg | \frac{1}{n}\sum_{i=1}^n I\{\Xvec_i \in B(\xvec, 1/n)\} -P\{\Xvec_1 \in B(\xvec, 1/n)\}\bigg | ~~~~~~~~~~~~~~~~~~~~~~~~~~
\vspace{-0.1in}
$$ $$~~~~~~~~~~~~~~~~~~~~~~~~~~~+ \frac{2K({\bf 0})}{h^d}P\{\Xvec_1 \in B(\xvec, 1/n)\} + \frac{2nK({\bf 0})\sqrt{2}}{n^2h^d}.
\vspace{-0.05in}
$$

Note that $I\{\Xvec_i\in B(\xvec, 1/n)\}$ are i.i.d. bounded random variables with expectation  $P\{\Xvec_1 \in B(\xvec, 1/n)\}$. So, the a.s. convergence of the first term follows from Bernstein's inequality. Since $P\{\Xvec_1 \in B(\xvec, 1/n)\} \le M_f n^{-d}$ (where $M_f = \sup_{\xvec} f(\xvec)<\infty$), the second term converges to $0$. The third term also converges to $0$ as $n \to \infty$. Therefore, we have $|\mbox{LSPD}(\xvec, F_n)-\mbox{LSPD}(\yvec,F_n)| \stackrel{a.s}{\rightarrow} 0$ as $n \to \infty$.

Similarly, one can prove that $|\mbox{LSPD}(\xvec, F)-\mbox{LSPD}(\yvec,F)| \stackrel{a.s}{\rightarrow} 0$  as $n \to \infty$. Note that in the arguments above, all bounds are free from $\xvec$ and $\yvec$. We have also proved that $\sup_{\yvec \in A_n} |\mbox{LSPD}(\yvec,F_n) -\mbox{LSPD}(\yvec, F)| \stackrel{a.s.}{\rightarrow} 0$ as $n \to \infty$. So, combining these results, we have
%$\sup_{\xvec \in B_n \setminus A_n} |LSPD(\xvec,F_n) -LSPD(\xvec,F)|\stackrel{a.s.}{\rightarrow} 0$ as $n$ tends to infinity.
$\sup_{\xvec \in B_n \setminus A_n}|\mbox{LSPD}_h(\xvec, F_n)-\mbox{LSPD}_h(\xvec, F)| \stackrel{a.s}{\rightarrow} 0~\mbox{as}~n \to \infty.$

%\vspace{.05 in}
(iii) Now, consider the region outside $B_n$ (i.e., $B_n^c$). First note that
\vspace{-0.05in}
{\small $$\sup_{\xvec \in B_n^c} |\mbox{LSPD}_h(\xvec,F_n)-\mbox{LSPD}(\xvec,F)| \le \sup_{\xvec \in B_n^c} \frac{1}{nh^d} \sum_{i=1}^{n}K(h^{-1}(\xvec-\Xvec_i)) + \sup_{\xvec \in B_n^c} h^{-d}E\left[K(h^{-1}(\xvec-\Xvec))\right].
\vspace{-0.05in}$$}
We will show that both of these terms become sufficiently small as $n \to \infty$.

%First note that for any  and $h>0$
Fix any $\epsilon >0$. We can choose two constants $M_1$ and $M_2$ such that $P(\|\Xvec\|\ge M_1)\le  h^d \epsilon/2K({\bf 0})$ and $K(\tvec)\le h^d \epsilon/2$ when $\|\tvec\| \ge M_2$. %Now, choose $n$ such that $n > 2\max\{M_1,M_2h\}$.
Now, one can check that
\vspace{-0.05in}
$$h^{-d}E\left[K(h^{-1}(\xvec-\Xvec))\right] \le h^{-d}E\left[K(h^{-1}(\xvec-\Xvec))I(\|\Xvec\|\le M_1)\right] + h^{-d}K({\bf 0})P(\|\Xvec\|> M_1).
\vspace{-0.05in}
$$
Note that if $\xvec \in B_n^c$ and $\|\Xvec\| \le M_1$, $h^{-1}\|\xvec - \Xvec\| \geq h^{-1} |\sqrt{d}n - M_1|$. Now, choose $n$ large enough such that $|\sqrt{d}n - M_1| \geq M_2h$, and this implies $K(h^{-1}(\xvec-\Xvec)) \leq h^d\epsilon/2$.
% for $\|\xvec\| \leq \sqrt{d}n$ and $\|\yvec\| > M_1$.
So, we get
\vspace{-0.05in}
$$h^{-d}E\left[K(h^{-1}(\xvec-\Xvec))\right] \le \epsilon/2 + h^{-d}K({\bf 0})P(\|\Xvec\|> M_1)\le \epsilon,~\mbox{and}$$
%\vspace{-0.1in}
$$~~~~~~~~~~~~~~~~~\frac{1}{nh^d} \sum_{i=1}^{n} K(h^{-1}(\xvec-\Xvec_i)) \le \epsilon/2 + h^{-d}K({\bf 0}) \frac{1}{n} \sum_{i=1}^{n} I(\|\Xvec_i\|>M_1)~~~~~~~~~~~~~~~~~~~~~~~~~$$ 
$$~~~~~~~~~~~~~~~~~~~~~~~~~~~~~~~~~~~~~~~~~~~~~~~~~\le \epsilon + h^{-d}K({\bf 0}) \left|\frac{1}{n} \sum_{i=1}^{n} I(\|\Xvec_i\|>M_1)-P(\|\Xvec\|> M_1)\right|.$$
The Glivenko-Cantelli theorem implies that the last term on the right hand side converges to $0$ as $n \rightarrow \infty$. So, we have
$\sup_{\xvec \in B_n^c}|\mbox{LSPD}_h(\xvec, F_n)-\mbox{LSPD}_h(\xvec, F)| \stackrel{a.s}{\rightarrow} 0~\mbox{as}~n \to \infty.$

Combining (i), (ii) and (iii), we now have $\sup_{\xvec}|\mbox{LSPD}_h(\xvec, F_n)-\mbox{LSPD}_h(\xvec, F)| \stackrel{a.s.}{\rightarrow} 0$ for any $h \in (0,1]$. 

For any fixed $h>1$, this a.s. convergence can be proved in a similar way. In that case, recall that the definition of LSPD does not involve the $h^d$ term in the denominator. \hfill $\Box$

%$\sup_{\xvec \in B_n^c} |LSPD(\xvec,F_n) -LSPD(\xvec,F)|\stackrel{a.s.}{\rightarrow} 0$ as $n$ tends to infinity.

%Therefore,
%$\sup_{\xvec \in B_n^c} h^{-d}E\left[K(h^{-1}(\xvec-\Xvec))\right] \le \epsilon/2+ \epsilon/2 =\epsilon.$
%Similarly, one can show that
%$\sup_{\xvec \in B_n^c} \frac{1}{nh^d} K(h^{-1}(\xvec-\Xvec_i)) \le \epsilon/2+ h^{-d}K({\bf 0})\frac{1}{n} \sum_{i=1}^{n} I(\|\Xvec_i\|>M_1)$\\
%$ ~~\hspace{0.3in}~~~~~~~~~~~~~~~~~~~~~~~~~~~~~~~~~~~~\le \epsilon + h^{-d}K({\bf 0})\left|\frac{1}{n} \sum_{i=1}^{n} I(\|\Xvec_i\|>M_1)-P(\|\Xvec\|> M_1)\right|$. Now, following the proof of part (i), we can show that $h^{-d}K({\bf 0})\left|\frac{1}{n} \sum_{i=1}^{n} I(\|\Xvec_i\|>M_1)-P(\|\Xvec\|> M_1)\right| \stackrel{a.s.}{\rightarrow} 0$ as $n \rightarrow \infty$.

\vspace{.025in}
\noindent
{\bf Remark 2:}  Following the proof of Theorem 3, it is easy to check that the a.s. convergence holds when $h$ diverges to infinity at any rate with $n$.

\vspace{.025in}
\noindent
{\bf Remark 3:} The result continues to hold when $h \rightarrow 0$ as well. However, %in that case,
%But, we need some additional conditions.
for the a.s. convergence in part (i), (more specifically, to use the Borel-Cantelli lemma), we require $nh^{2d}/\log n \rightarrow \infty$ as $n \rightarrow \infty$. 
In part (iii), we need $M_1$ and $M_2$ to vary with $n$. Assume the first moment of $f$ to be finite, and $\int \|\tvec\| K(\tvec) d\tvec < \infty$ (which implies $\|\tvec\| K(\tvec) \to 0$ as $\|\tvec\| \to \infty$). 
%Given finiteness of the first moments, and 
Also assume that $nh^{2d}/\log n \rightarrow \infty$ as $n \rightarrow \infty$. 
We can now choose $M_1=M_2=\sqrt{n}$ to ensure that both $P(\|\Xvec\|\ge M_1)\le  h^d \epsilon/2K({\bf 0})$ and $K(\tvec)\le h^d \epsilon/2$ for $\|\tvec\| \ge M_2$ hold for sufficiently large $n$.

%$M_1$ and $M_2$ varies with $h$, and in order to choose $n/2>\max\{M_1,M_2h\}$, we need existence of the first moments of $f$ and $K$.? %Theorem 2 holds for the empirical version of LSPD under these additional assumptions.

%Suppose that $M_f = \sup_{\xvec} f(\xvec)$, $M_f^{\prime} = \sup_{\xvec} \|\nabla f(\xvec)\|$, $M_K = \sup_{\xvec} K(\xvec)$ and $M_K^{\prime} = \sup_{\xvec} \|\nabla K(\xvec)\|$ are all finite. Also assume that $I_K=\int \|\tvec\|K(\tvec)d\tvec$ is finite.
%  $h=n^{-\beta}$ with $1/4(d+1) < \beta < 1/2d$, we have
%For $h \rightarrow 0$ in such a way that $nh^{2d} \rightarrow \infty$ and $\log n/nh^{2d} \rightarrow 0$ as $n \rightarrow \infty$, we have
%$$\sup_{\xvec} |\mbox{LSPD}_{h}(\xvec,F_{n}) - f(\xvec)| \stackrel{P}{\rightarrow} 0~\mbox{as}~n \rightarrow \infty.$$

\vspace{.05in}
\noindent
{\bf Proof of {Theorem 4 (a)} :}
Consider two independent random vectors $\Xvec=(X^{(1)}, \ldots$, $X^{(d)})^T \sim F_j$ and $\Xvec_{1}=(X_1^{(1)}, \ldots, X_1^{(d)})^T \sim F_j$, where $1 \leq j \leq J$.
% Since
% $$\frac{\|\Xvec-\Xvec_{1}\|^2}{{d}}$$
% $$= d^{-1} \sum_{k=1}^d (X^{(k)})^2 + d^{-1} \sum_{k=1}^d (X_{1}^{(k)})^2 - 2d^{-1} \sum_{k=1}^d X^{(k)}X_{1}^{(k)},$$
It follows from (C1) and (C2) that
${\|\Xvec-\Xvec_{1}\|}/{\sqrt{d}} \stackrel{a.s.}{\rightarrow} \sqrt{2\sigma_j^2}~\mbox{as}~d \rightarrow \infty.$
% where $\sigma_i=a_i-b_{ii}$.
So, for almost every realization $\xvec$ of $\Xvec \sim F_j$,
\vspace{-0.05in}
\begin{equation}
{\|\xvec-\Xvec_{1}\|}/{\sqrt{d}} \stackrel{a.s.}{\rightarrow} \sqrt{2\sigma_i^2}~\mbox{as}~d \rightarrow \infty.
\vspace{-0.05in}
\end{equation}
Next, consider two independent random vectors $\Xvec \sim F_j$ and $\Xvec_{1} \sim F_i$ for $1 \leq i \neq j \leq J$. Using (C1) and (C2), we get
$
{\|\Xvec - \Xvec_{1}\|}/{\sqrt{d}} \stackrel{a.s.}{\rightarrow} \sqrt{\sigma_{j}^2+\sigma_{i}^2+\nu_{ji}}~\mbox{as}~d \rightarrow \infty.
$
% where $\nu_{ij} = b_{ii}-2b_{ij}+b_{jj}$.
Consequently, for almost every realization $\xvec$ of $\Xvec \sim F_j$
\vspace{-0.05in}
\begin{equation}
{\|\xvec - \Xvec_{1}\|}/{\sqrt{d}} \stackrel{a.s.}{\rightarrow} \sqrt{\sigma_{j}^2+\sigma_{i}^2+\nu_{ji}}~\mbox{as}~d \rightarrow \infty,
\vspace{-0.05in}
\end{equation}

% \noindent
Let us next consider $\langle \xvec - \Xvec_{1}, \xvec - \Xvec_{2} \rangle$, where $\Xvec \sim F_j$,  $\Xvec_{1}, \Xvec_{2} \sim F_i$ are independent random vectors, and $\langle \cdot, \cdot \rangle$ denotes the inner product in $\mathbb{R}^d$. Therefore, for almost every realization $\xvec$ of $\Xvec$, arguments similar to those used in (8) and (9) yield
\begin{eqnarray}
\frac{\langle \xvec - \Xvec_{1}, \xvec - \Xvec_{2} \rangle}{d} \stackrel{a.s.}{\rightarrow} {\sigma_j^2}~\mbox{as}~d \rightarrow \infty~\mbox{if}~1 \leq i=j \leq J,~\mbox{and}\\
\frac{\langle \xvec - \Xvec_{1}, \xvec - \Xvec_{2} \rangle}{d} \stackrel{a.s.}{\rightarrow} {\sigma_j^2+\nu_{ji}}~\mbox{as}~d \rightarrow \infty~\mbox{if}~1 \leq i \neq j \leq J.
\end{eqnarray}
% \noindent
Observe now that
$\| E_{F_j}[u(\xvec - \Xvec)] \|^2 = \langle E_{F_j}[u(\xvec - \Xvec_1)], E_{F_j}[u(\xvec - \Xvec_{2})] \rangle = E_{F_j} \{ \langle u(\xvec - \Xvec_{1}), u(\xvec - \Xvec_{2}) \rangle \},$
% $$ \hspace{1.2in} = E_{F_i} \{ \langle u(\Xvec - \Xvec_{1}), u(\Xvec - \Xvec_{2}) \rangle | \Xvec = \xvec \}. $$
where $\Xvec_1, \Xvec_2 \sim F_j$ are independent random vectors for $1 \leq j \leq J$.
% , and
% $$ \langle u(\xvec - \Xvec_{1}), u(\xvec - \Xvec_{2}) \rangle = \bigg \langle \frac{\xvec - \Xvec_{1}}{\|\xvec - \Xvec_{1}\|}, \frac{\xvec - \Xvec_{2}}{\|\xvec - \Xvec_{2}\|} \bigg \rangle.$$

% \noindent
Since here we are dealing with expectations of random vectors with bounded norm, a simple application of Dominated Convergence Theorem implies that for almost every realization $\xvec$ of $\Xvec \sim F_j$  $(1 \leq i \leq J)$, as $d \rightarrow \infty$,
\begin{eqnarray}
{\mbox{SPD}(\xvec,F_{j})} \stackrel{a.s.}{\rightarrow} 1-\sqrt{\frac{1}{2}}
~\mbox{and}~{\mbox{SPD}(\xvec,F_{i})} \stackrel{a.s.}{\rightarrow} 1-\sqrt{\frac{\sigma_j^2+\nu_{ji}}{\sigma_j^2+\sigma_i^2+\nu_{ji}}} ~\mbox{for}~ i \neq j.
\end{eqnarray}
% \noindent
Therefore, for $\Xvec \sim F_j$, we get
$z(\Xvec)=({\mbox{SPD}(\Xvec, F_{1}}), \ldots, {\mbox{SPD}(\Xvec, F_{J}}))^T \stackrel{a.s.}{\rightarrow} \cvec_{j},$
as $d \rightarrow \infty$.%, where $\cvec_i$ is as defined in {Theorem 3 (a)}.
\hfill $\Box$
% This completes the proof.

\vspace{.05in}
\noindent
{\bf Proof of {Theorem 4 (b)} :} Recall that for $h > 1$,
${\mbox{LSPD}}_h(\xvec,F) =  E_F[h^d{K}_h(\tvec)] - \| E_F[h^dK_h(\tvec) u(\tvec)] \|$, and since we have assumed $\Xvec$s to be standardized, here we have $h^dK_h(\tvec) = K((\xvec - \Xvec)/h) = g(\|\xvec - \Xvec\|/h)$.
Let $\Xvec \sim F$ and $\Xvec_i \sim F_i$ where $1 \leq i \leq J$.
Then, using (8) and (9) above, and the continuity of $g$, for almost every realization $\xvec$ of $\Xvec \sim F_j$, one gets
$$g \biggl(\frac{\|\xvec - \Xvec_i\|}{\sqrt{d}} \frac{\sqrt{d}}{h} \biggr) \stackrel{a.s.}{\rightarrow} g(0)~\mbox{or}~g(e_{ji}A),$$
depending on whether ${\sqrt{d}}/{h} \rightarrow 0 ~\mbox{or}~ A$, for almost every realization $\xvec$ of $\Xvec \sim F_j$.
% , and $e_{ij}$ is defined as $\sqrt{2\sigma_i^2}$ if $i=j$, and $\sqrt{\sigma_i^2+\sigma_j^2+\nu_{ij}}$ if $i \neq j$.
The proof now follows from a simple application of Dominated Convergence Theorem. %be completed using arguments that are very similar to the arguments given in the proof of {Theorem 3 (a)}.
\hfill $\Box$

\vspace{.05in}
\noindent
{\bf Proof of {Theorem 4 (c)} :} Since $g(s) \rightarrow 0$ as $s \rightarrow \infty$, using the same argument as used in the proof of Theorem 3(b),
for $\Xvec_i \sim F_i$ and almost every realization $\xvec$ of $\Xvec \sim F_j$, we have
$$g \biggl(\frac{\|\xvec - \Xvec_i\|}{\sqrt{d}} \frac{\sqrt{d}}{h} \biggr) \stackrel{a.s.}{\rightarrow} 0 ~~\mbox{as}~ {\sqrt{d}}/{h} \rightarrow \infty. $$
The proof now follows from a simple application of Dominated Convergence Theorem. \hfill $\Box$

\vspace{.05in}
\noindent
{\bf Lemma 2 :} Recall $\cvec_j$ and $\cvec^{\prime}_j$ for $1 \leq j \leq J$ defined in Theorem 3 (a) and (b), respectively. For any $1 \leq j \neq i \leq J$, $\cvec_j = \cvec_i$ if and only if $\sigma_j = \sigma_i$ and $\nu_{ji} = \nu_{ij} = 0$.
Similarly, $\cvec_j^{\prime} = \cvec_i^{\prime}$ if and only if $\sigma_j = \sigma_i$ and $\nu_{ji} = \nu_{ij} = 0$.

\vspace{.05in}
\noindent
{\bf Proof of Lemma 2 :} The `if' part is easy to check in both cases.  So, it is enough to prove the `only if' part and that too for the case of $J=2$. Note that if $\cvec_1=(c_{11},c_{12})^{T}$ and $\cvec_2=(c_{21},c_{22})^{T}$ are equal, we have
$${\frac{\sigma_1^2+\nu_{12}}{\sigma_1^2+\sigma_2^2+\nu_{12}}}=1/2 ~~\mbox{and}~~ {\frac{\sigma_2^2+\nu_{12}}{\sigma_1^2+\sigma_2^2+\nu_{12}}}=1/2.$$
These two equations hold simultaneously only if $\sigma_1^2=\sigma_2^2$ and $\nu_{12}(=\nu_{21})=0$.

% \noindent
% For the $\cvec^{\prime}_i$s also we need to show that $\cvec^{\prime}_1 = \cvec^{\prime}_2$ implies $\sigma_1^2 = \sigma_2^2$ and $\nu_{12} = 0$.
Now consider the case $\cvec_1^{\prime}=\cvec_2^{\prime}$.
% So, we assume that $\cvec_{1}^{\prime} = \cvec_{2}^{\prime}$.
Recall that $c_{11}^{\prime}=g(A\sqrt{2}\sigma_1) c_{11}$, $c_{22}^{\prime}=g(A\sqrt{2}\sigma_2) c_{22}$,  $c_{12}^{\prime}=g(A\sqrt{\sigma_1^2+\sigma_2^2+\nu_{12}}) c_{12}$ and $=c_{21}^{\prime}=g(A\sqrt{\sigma_2^2+\sigma_1^2+\nu_{21}}) c_{21}$.
% We now present a proof by the method of contradiction.
If possible, assume that $\sigma_1 > \sigma_2$. This implies that
$A \sqrt{\sigma_1^2+\sigma_2^2+\nu_{12}} > A \sqrt{2}\sigma_1$ and hence
\begin{equation}
g(A\sqrt{2}\sigma_1) > g(A\sqrt{\sigma_1^2+\sigma_2^2+\nu_{12}})~~~ \mbox{(since
$g$ is monotonically decreasing)}.
\end{equation}
Also, if $\sigma_1 > \sigma_2$, we must have
\begin{equation}
1/2 < \frac{\sigma_1^2}{\sigma_1^2+\sigma_2^2} < \frac{\sigma_1^2+\nu_{12}}{\sigma_1^2+\sigma_2^2+\nu_{12}} < 1 \Leftrightarrow 1 - \sqrt{1/2} > 1 - \sqrt{\frac{\sigma_1^2+\nu_{12}}{\sigma_1^2+\sigma_2^2+\nu_{12}}}.
\end{equation}
Combining (13) and (14), we have $c^{\prime}_{11} > c^{\prime}_{21}$, and this implies $\cvec^{\prime}_1 \neq \cvec^{\prime}_2$.
Similarly, if $\sigma_1 < \sigma_2$, we get $c^{\prime}_{12} > c^{\prime}_{22}$ and hence $\cvec^{\prime}_1 \neq \cvec^{\prime}_2$.
% So, we conclude that $\sigma_1 = \sigma_2$.
Again, if $\sigma_1 = \sigma_2$ but $\nu_{12} = \nu_{21} > 0$, similar arguments lead to $\cvec^{\prime}_1 \neq \cvec^{\prime}_2$ . This completes the proof of the lemma. \hfill $\Box$

\vspace{-.15in}
%\newpage

\end{document}